\documentclass{aa} 

\usepackage{aabib99}
\usepackage{graphics} \usepackage{psfig}
\usepackage{rotating} \usepackage{times}
\newcommand{\gapprox}{\,\rlap{\lower 2.5pt 
\hbox{$\sim$}}\raise 1.5pt\hbox{$>$}\,}
\newcommand{\lapprox}{\,\rlap{\lower 2.5pt 
\hbox{$\sim$}}\raise 1.5pt\hbox{$<$}\,}

\begin{document}

   \thesaurus{11     
              (11.19.4;  
               11.19.5;  
               11.09.2;  
               08.16.4;  
               08.05.3)} 
   \titlerunning{The AGB phase transition outside the Local Group}

   \title{The AGB phase-transition outside the Local Group: 
    K-band observations of young star clusters in NGC 7252}

   \author{C. Maraston \inst{1}
           \and
           M. Kissler-Patig \inst{2}
	   \and
	   J.P. Brodie \inst{3}
	   \and
	   P. Barmby \inst{4}
	   \and
           J.P. Huchra \inst{4} 
            }

   \offprints{C. Maraston}

   \institute{Universit\"ats-Sternwarte der Ludwig-Maximilians-Universit\"at,
	       Scheinerstr.~1, 81679 M\"unchen, Germany\\
	       email: maraston@usm.uni-muenchen.de
          \and
	       European Southern Observatory,
		Karl-Schwarzschild-Str.~2, 85748 Garching, Germany\\
              email: mkissler@eso.org
	  \and
	  	University of California Observatories / Lick Observatory,
		University of California, Santa Cruz, CA 95064, USA\\
              email: brodie@ucolick.org
          \and 
	        Havard-Smithsonian Center for Astrophysics,
		60 Garden St., Cambridge, MA 02138, USA\\
              email: pbarmby@cfa.harvard.edu, huchra@cfa.harvard.edu
             }

   \date{Received September 2, 2000; accepted January 29, 2001}

   \maketitle

   \begin{abstract}

We have extended the study of the young star clusters observed in the
merger remnant galaxy NGC 7252 by obtaining $K$ band photometry for
these clusters.  Our $K$ band data significantly complement the optical
photometry and spectroscopy in the literature: $K$ band data are
fundamental to study the Asymptotic Giant Branch (AGB) population of
these clusters, since the AGB {\it phase transition} (occuring between
the age of $\sim 200$ Myr and $\sim$ 1 Gyr) causes abrupt changes in
the near-infrared luminosity of the clusters while producing only
small changes in the optical. Therefore, the e.g.~V$-$K colour is
ideal to study this evolutionary phase of stellar populations.

For the present analysis we present models for Simple Stellar
Populations which include the contribution of the AGB stellar phase,
calibrated with the young and intermediate age star clusters of the
Magellanic Clouds. The comparison with the colour distribution of the
NGC 7252 star clusters shows that they are indeed intermediate age
clusters undergoing the AGB phase transition. The AGB phase transition
is observed for the first time outside the Local Group.

Most of the studied clusters span the very narrow age range
$300\,$--$\,500$~Myr, and likely have metallicities
$0.5\,$--$\,1~Z_{\odot}$.  A very important exception is the cluster
W32, which has already completed its AGB epoch, its colours being
consistent with an age of $\sim 1\,$--$\,2$~Gyr. This impacts on the
duration of the merger-induced starburst.

The strengths of the magnesium and iron lines in the spectrum of the
best observed cluster W3, and in the spectrum of the diffuse central
light of NGC 7252, do not show an overabundance in $\alpha$-elements,
in contrast to the bulk stellar population of elliptical galaxies.


  \end{abstract}


\section{Introduction}

Over the last decade, many systems of luminous star clusters have been
discovered in merging and merger remnant galaxies (Lutz 1991;
Holtzmann et al. 1992; see also Schweizer 1998 for a review). Many of
these appear to be globular clusters based on their compactness (small
effective radius). They are thought to be young systems ($t<1$~Gyr) on
the basis of their blue broadband colours and high luminosities. The
low ages support the view that they formed during the past merger. The
typical duration of the merger-induced star formation episode is about
a few hundred million years \cite{MH96}.  These newly formed globular
clusters provide the opportunity to study the process of globular
cluster formation, and the early phases of the evolution of their
stellar populations.

 From the theoretical side, a stellar population aging to $t\sim
200\,$--$\,400$~Myr (intermediate-age range) abruptly develops a well
populated Asymptotic Giant Branch (AGB), and consequently displays
very red colours. When the age exceeds $\sim 1$~Gyr, the contribution
of AGB stars sharply decreases. This very short epoch is referred to
as the AGB phase-transition (Renzini 1981). The observational
counterpart of this theoretical prediction is represented by the
intermediate-age GCs in the Magellanic Clouds, which span a large
range in V$-$K (from~$\sim 1$ to $\sim 3$ mag), as a consequence of
the AGB phase-transition (see e.g. Persson et al. 1983, Frogel, Mould
\& Blanco 1990). The Magellanic Clouds are the only sites in which the
AGB phase transition has been observed so far.  Recent mergers are the
natural places to further investigate the occurrence of such a
phenomenon.

The galaxy NGC 7252 is a prototypical remnant of two merged disk-like
galaxies (Fritze-v Alvensleben \& Gerhard 1994a,b; Schweizer 1998),
containing very bright young globular clusters (Schweizer 1982;
Whitmore et al. 1993). A deep optical photometric study with {\em HST} was
carried out by Miller et al. (1997), who conveniently divided the star
clusters of NGC 7252 into two populations. The ``outer'' population
contains objects with projected galactocentric distances between
6\arcsec\ and $\sim$ 120\arcsec , the ``inner'' population contains
objects within 6\arcsec\ of the nucleus.  By analyzing the object
colours and using a reddening-free parameter, Miller et al. (1997)
conclude that the inner sample mainly consists of very young OB
associations, while the outer sample is composed by bright blue
(young) globular clusters and by fainter reddish (old) globular
clusters.  Schweizer \& Seitzer (1998) performed a spectroscopic study
of a sub-sample of globular clusters belonging to the outer blue
population.  Ages for these clusters were determined via comparison of
the strong observed Balmer lines with various sets of models, and
found to be in the narrow range $400\,$--$\,600$~Myr. The ages determined from
$V,I$ optical photometry were found to be 20 \% smaller than the
spectroscopic ones.

If the outer blue population of NGC 7252 contains intermediate-age
clusters, the evidence for AGB phase transition is expected among
these clusters. With this in mind, we obtained $K$ magnitudes for a
sample of globular clusters belonging to the outer blue sample. The
$K$ photometry, not included in previous studies, was motivated by the
fact that AGB stars mainly emit in the infrared (see e.g. Persson et
al. 1983). Complementing with published optical photometry, we analyze
the data in the two-colour diagram V$-$K vs. B$-$V, using Simple
Stellar Population models {\it calibrated} with the intermediate-age
star clusters in Magellanic Clouds (Maraston 1998).  The spectroscopy
by Schweizer \& Seitzer (1998) is used here to further constrain the
cluster metallicities, and to investigate the element abundances.  The
galaxy NGC 7252 is the only remnant so far to have been modelled in detail
using $N$-body simulations (Fritze-v Alvensleben \& Gerhard 1994b;
Hibbard \& Mihos 1995).  The derived age distribution of our cluster
sample can therefore be compared with the predictions of these
simulations.

The paper is organized as follows. Section~2 describes the observations
and the data reduction. The index measurements on published
spectroscopy are also presented here.  The modelling of intermediate
ages SSPs is presented in Section~3. In Section 4, we discuss the photometric 
ages for individual clusters and the comparison with the
spectroscopically-derived values, while Section~5 deals with
metallicities and abundance ratios for the star clusters and the
diffuse galaxy light. Results are discussed in Section~6 and conclusions
are drawn in Section~7.

\section{The data}

\subsection{Observations and data reduction}

The $K$-band data were obtained with the Near-Infrared Camera (NIRC,
Matthews \& Soifer 1994) on the Keck I telescope on the night
of June 20, 1999. The camera operates with a 256 $\times$ 256 pixels InSb
detector, with a pixel size of 30$\mu$m corresponding to 0.15\arcsec\ on
the sky. The total field of view is 38\arcsec $\times$ 38\arcsec .

We used a $K_{\rm s}$ filter for all observations (see Persson et al. 1998
for the transmission curve). We tested various flatfields and finally
used one computed from a number of random sky observations done
shortly before observing NGC 7252. These were dark subtracted,
averaged using a sigma clipping algorithm and normalized. The
flatfield was then applied to all science and calibration data.

\begin{figure*}
\psfig{figure=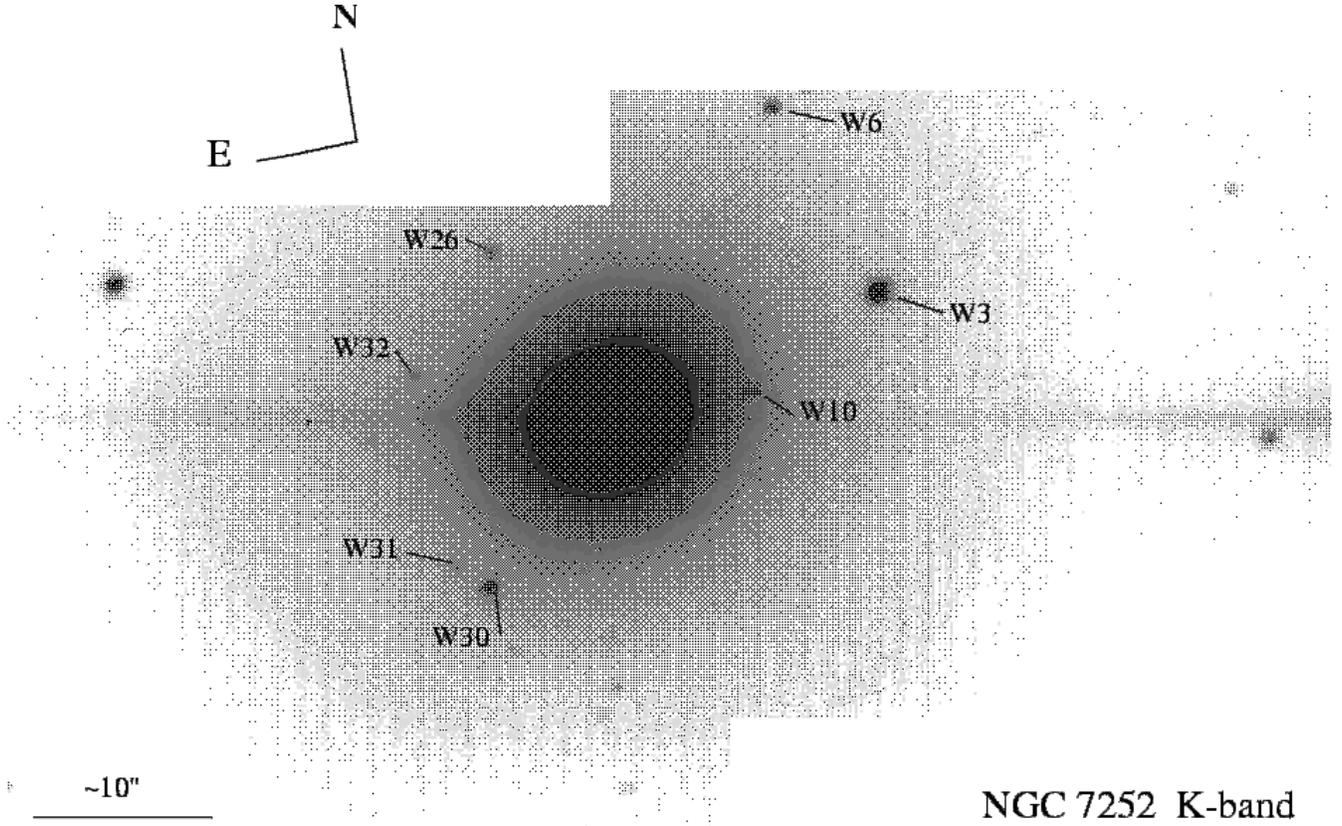,width=\linewidth}
\caption{The composite image of the east and west field. Some clusters
are marked as reference, however the contrast chosen for display does not 
allow us to show all detected clusters.}
\label{image7252}
\end{figure*}

Two adjacent, slightly overlapping, fields were centered on NGC 7252.
The east-west coverage is $\simeq$ 70 \arcsec , the north-south
coverage is 38\arcsec . The observations were carried out in 300s
blocks: each block was split into 100s (5 coadds of 20s) on the east
field, 100s (5 coadds of 20s) on the west field, and 100s on the sky
(250\arcsec\ north), the latter being divided into 5 dithered
exposures (5 coadds of 4s each).  These 5 sky exposures were averaged
with a sigma clipping algorithm to create a blank sky field. This sky
field (of 20s effective exposure time) was normalized to 100s
effective exposure time and subtracted from both the east and the west
field of the same block.  The individual images of the east and the
west field, obtained this way, were then respectively averaged to
compute the final images.  The total integration time on each NGC 7252
field was 1300s.  The seeing on the final images, as measured from the
FWHM of the objects, was around 0.5\arcsec .  A composite image of
both fields in shown in Fig.\ref{image7252}.
\footnote{A full version of the image in Fig.1 can be retrieved on 
 http://usm.uni-muenchen.de/~maraston.}

\subsection{$K$-Photometry}

Seven standards from Persson et al. (1998) were observed throughout
the night. Most standards were observed more than once at different
airmasses. Each standard observation is composed of nine exposures (4
coadds of 1s) in a box pattern. The resulting 144 standard
measurements were used to derive a zero point and extinction
coefficient; the colour term was assumed to be negligible. We obtained
the following relation:\\ $K_{\rm s} = 24.797 (\pm0.021) + (m_{inst} -
0.046 (\pm 0.005) \cdot x)$,\\ where $m_{inst}$ is the measured
magnitude of an object (normalized to 1s exposure) and $x$ is the airmass.

In order to remove any steep background gradients and to perform the
photometry on a homogeneous ``sky'', we modelled the galaxy with a
31$\times$31 pixels median filter and subtracted it from the original
images.

The photometry was performed on the galaxy-subtracted images with 
{\tt SExtractor} \cite{sex96}. The
detection parameters were set to 5 adjacent pixels 1.5 $\sigma$ over
the local background (as defined in a 32$\times$32 pixels box). The
photometry was computed in different apertures, and finally the
``automatic aperture magnitudes'' (see Bertin \& Arnouts 1996) were used for
consistency with the standard star measurements. The K$_{\rm s}$ magnitudes
were corrected for a Galactic foreground extinction of $E(B-V)=0.012$
(for consistency with Miller et al. 1997, see below), using the
relation $A_K=0.36\cdot E(B-V)$ (Fitzpatrick 1999).

\begin{table}
\caption[]{K$_{\rm s}$ magnitudes of young star clusters in NGC 7252. Note
that the magnitudes in this table are \emph{not} corrected for
Galactic foreground extinction. The (V$-$K) colour, however, is
corrected for a Galactic foreground extinction of $E(B-V)=0.012$.}
\label{kmags}
\begin{tabular}{l c c}
\hline
Name & K$_{\rm s}$ & (V$-$K$_s$)$_0$ \\
\hline
W3   & 15.39 $\pm$ 0.03 & 2.45 $\pm$ 0.04 \\ 
W6   & 17.37 $\pm$ 0.03 & 2.27 $\pm$ 0.04 \\
W19  & 20.15 $\pm$ 0.14 & 2.30 $\pm$ 0.17 \\
W22  & 19.13 $\pm$ 0.10 & 2.48 $\pm$ 0.10 \\
W24  & 20.22 $\pm$ 0.15 & 2.47 $\pm$ 0.15 \\
W25  & 19.94 $\pm$ 0.13 & 2.53 $\pm$ 0.13 \\
W26  & 18.13 $\pm$ 0.03 & 2.26 $\pm$ 0.04 \\
W27  & 20.11 $\pm$ 0.09 & 1.98 $\pm$ 0.09 \\
W30  & 17.06 $\pm$ 0.03 & 2.40 $\pm$ 0.04 \\
W31  & 19.11 $\pm$ 0.06 & 1.96 $\pm$ 0.06 \\
W32  & 18.79 $\pm$ 0.05 & 2.75 $\pm$ 0.05 \\
S105 & 19.06 $\pm$ 0.06 &        ....     \\
S114 & 18.99 $\pm$ 0.08 & 2.24 $\pm$ 0.09 \\
\hline
\end{tabular}
\end{table}

The final list of K$_{\rm s}$ magnitudes (uncorrected for extinction)
are given in Table 1. We followed the naming convention of Schweizer
\& Seitzer (1998), mostly referring to the list of Whitmore et
al. (1993).

\subsection{Previous optical and UV photometry}

The $UV$ and optical photometry is taken from Miller et al. (1997),
except for object S114 for which we used the photometry given by
Schweizer \& Seitzer (1998), and for object W19, for which we used the
photometry given in Whitmore et al. (1993). There is no optical
photometry available for object S105.

\begin{table}
\caption[]{Optical magnitudes of young star clusters in NGC 7252 (see
references in the text). All values are corrected for a Galactic foreground
extinction of $E(B-V)=0.012$.}
\label{opticalmags}
\begin{tabular}{l c c c c}
\hline
Name & V$_0$  & (U$-$B)$_0$  & (B$-$V)$_0$ & (V$-$I)$_0$ \\
\hline
W3   & 17.84 $\pm$ 0.02 & 0.28 $\pm$ 0.06 & 0.45 $\pm$ 0.04 & 0.64 $\pm$ 0.04 \\
W6   & 19.64 $\pm$ 0.02 & 0.31 $\pm$ 0.06 & 0.44 $\pm$ 0.04 & 0.64 $\pm$ 0.03 \\
W19  & 22.45 $\pm$ 0.10 &    ....         &   ....          & 0.79 $\pm$ 0.12 \\
W22  & 21.61 $\pm$ 0.02 & 0.20 $\pm$ 0.08 & 0.44 $\pm$ 0.04 & 0.68 $\pm$ 0.04 \\
W24  & 22.69 $\pm$ 0.03 & 0.35 $\pm$ 0.17 & 0.41 $\pm$ 0.06 & 0.69 $\pm$ 0.05 \\
W25  & 22.47 $\pm$ 0.02 & 0.57 $\pm$ 0.16 & 0.36 $\pm$ 0.05 & 0.63 $\pm$ 0.04 \\
W26  & 20.39 $\pm$ 0.02 & 0.31 $\pm$ 0.06 & 0.53 $\pm$ 0.04 & 0.68 $\pm$ 0.03 \\
W27  & 22.09 $\pm$ 0.02 & 0.37 $\pm$ 0.10 & 0.38 $\pm$ 0.04 & 0.69 $\pm$ 0.04 \\
W30  & 19.46 $\pm$ 0.02 & 0.26 $\pm$ 0.06 & 0.41 $\pm$ 0.04 & 0.63 $\pm$ 0.03 \\
W31  & 21.07 $\pm$ 0.02 & 0.27 $\pm$ 0.07 & 0.29 $\pm$ 0.04 & 0.54 $\pm$ 0.03 \\
W32  & 21.54 $\pm$ 0.02 & 0.97 $\pm$ 0.18 & 0.80 $\pm$ 0.04 & 1.08 $\pm$ 0.04 \\
S105 &      ....        &     ....        &      ....       &   ....          \\
S114 & 21.23 $\pm$ 0.03 & 0.32 $\pm$ 0.09 & 0.42 $\pm$ 0.06 & 0.62 $\pm$ 0.05 \\
\hline
\end{tabular}
\end{table}

The magnitudes used in our work are given in Table~\ref{opticalmags}.
All magnitudes are corrected for Galactic foreground extinction adopting
$E(B-V)=0.012$, following Miller et al. (1997).

We note that Miller et al. (1997) give a warning to potential users of
their $U$-band photometry for which they note, among other problems, a
0.16 magnitude offset when compared to ground-based work. 
Therefore we refrain from using the (U$-$B) colours in our photometric
analysis. The $B$, $V$ and $I$ photometry is unaffected by problems.

\subsection{Spectroscopy} 

To date, the only database of spectroscopic observations of young star
clusters in NGC 7252 has been provided by Schweizer \& Seitzer (1998).

\begin{figure}
\resizebox{\hsize}{!}{\includegraphics{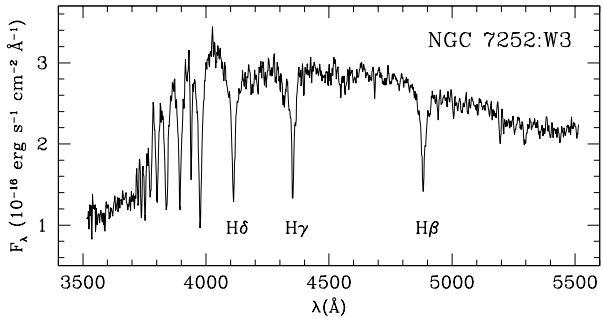}}
\caption{Ultraviolet-to-visual spectrum of cluster W3 in NGC 7252,
from Schweizer \& Seitzer (1998). Note the strong Balmer lines in absorption.}
\label{w3}
\end{figure}

The authors present ultraviolet-to-visual spectra for eight objects of
the outer blue population, namely W3, W6, W26, W30, W31, S105, S114,
S101. With the exception of S101, whose spectrum is dominated by
emission lines due to a surrounding giant H {\rm II} region, all the
spectra are characterized by strong Balmer lines (H$\beta\sim
6\,$--$\,13$~\AA) in absorption (see Fig.~\ref{w3}).  These features
are used by Schweizer \& Seitzer (1998) to determine the ages of the
clusters, by comparison with a version of the models by Bruzual \&
Charlot (1996, quoted as {\it private communication}) computed with
the needed high resolution of the observed spectra ($\lambda\la
2$~\AA). This restricts their comparisons to solar metallicity models,
because of the lack of high-resolution model spectra for other
metallicities.  In the next Section we present new measurements of the
Balmer lines for the cluster sample of Schweizer \& Seitzer
(1998). These will be compared with our model spectra, on which the
ages determined from photometry are based (see Section 4).

\subsection{The new index measurements} 
\label{newspe}

The cluster spectra by Schweizer \& Seitzer (1998, courtesy of F. Schweizer, 
{\it private communication}) are used as a complement to our photometric 
analysis (see Section 4).  In order to allow a sensible comparison, observed
and model spectra must have the same resolution, and the strengths of
relevant absorption features must be measured exactly in the same way
on the observed and the synthetic spectrum. Therefore the spectra of
Schweizer \& Seitzer (1998) have been broadened to match the resolution of 
20 \AA~(in the visual region) of the model spectra.

The equivalent widths (EWs) of the Balmer lines H$\beta$, H$\gamma$
and H$\delta$ are then measured adopting the definitions given in
Brodie et al. (1998, LS indices, their Table~3), designed to study
young stellar populations. The bandpass definitions are similar to the
ones given in Bruzual \& Charlot (1993), but with a different choice
for the continuum. The flux in the continuum for the LS indices is
determined as the average flux in the continuum bandpasses while it
coincides with the flux at the end points of the continuum bandpasses
in Bruzual \& Charlot (1993). Therefore the LS indices are better
suited to spectra with low S/N. They agree well with the Bruzual \&
Charlot ones for spectra with high signal-to-noise (see Brodie et
al. 1998 for further details). 
We chose both for the modelled spectra and the observed ones to measure
the EWs by centering the index definitions on the absorption feature,
in order to circumvent possible uncertainties in the rest frame
corrections of the spectra. Minimal corrections are necessary and do not
significantly affect the EW measurements.

\begin{table}
\caption[]{Equivalent widths of Balmer lines H$\beta$, H$\gamma$
and H$\delta$, for star clusters in NGC 7252 (spectra from
F. Schweizer). An indicative value of the
signal-to-noise ratio per \AA\ is given in the last column.}
\label{indices}
\begin{tabular}{l c c c c l}
\hline Name & H$\beta$ (\AA)& H$\gamma$ (\AA)& H$\delta$ (\AA) & S/N \\ 
\hline 
W3 & 10.1 $\pm0.3$ & 9.1 $\pm0.1$ & 12.2 $\pm0.5$ & $\sim$ 20 \\ 
W6 & 9.1 $\pm0.6$ & 8.5 $\pm0.6$ & 11.7 $\pm0.3$ & $\sim$ 8 \\
W26 & 7.3 $\pm2.9$ & 8.3 $\pm2.2$ & 13.8 $\pm1.5$ & $\sim$ 5 \\ 
W30 & 11.7 $\pm1.0$ & 9.3 $\pm1.2$ & 12.0 $\pm0.3$ & $\sim$ 8 \\ 
W31 & 6.2 $\pm3.5$ & 6.9 $\pm3.8$ & 12.5 $\pm5.8$ & $\sim$ 3 \\ 
S105 & 9.0 $\pm2.0$ & 12.7 $\pm3.0$ & 4.7 $\pm2.7$ & $\sim$ 3 \\ 
S114 & 5.4 $\pm1.1$ & 8.2 $\pm1.6$ & 9.5 $\pm5.4$ & $\sim$ 3 \\ 
\hline
\end{tabular}
\end{table}

The resulting Balmer line EWs are given in Table~\ref{indices}. The
quoted errors were taken from Schweizer \& Seitzer (1998). Since we
did not have the raw spectra, we were not able to estimate errors
based on photon noise.  A reference value of the signal-to-noise
ratio, as determined by a visual inspection of the spectra, is listed
in the last column.

It is worth mentioning that for strong, well-defined absorption
features the broadening procedure does not imply loss of
information. For comparison, we compute the same EWs on the original
($\sim 2$~\AA~resolution) spectra from Schweizer \& Seitzer (1998). On
average, the higher resolution implies larger EWs, the difference
being very small ($\sim 0.08$~\AA) for high S/N spectra (like W3), and
as large as $\sim 4$~\AA~for poorly observed spectra (like S105).

The EWs in Table~\ref{indices} can be compared to those in Table~3 of
Schweizer \& Seitzer (1998), bearing in mind that the latter refer to
different bandpasses and continuum levels, and are measured on the
$\sim 2$~\AA~resolution spectra. The comparison shows that the
relative strenghts of Balmer lines for a specific cluster, are the
same as in Schweizer \& Seitzer (1998).  Also, the relative strenghts
of Balmer lines of the cluster sample are consistent, with W3 and W30
exhibiting the strongest Balmer lines. The values in our
Table~\ref{indices} will be used in Section 4.3 to derive
spectroscopic cluster ages.

\section{The AGB phase transition in intermediate-age stellar populations}
\label{models}

The first development in a stellar population of stars with degenerate
carbon-oxygen cores is the appearance of an extended hydrogen and
helium shell-burning phase close to the Hayashii line, called the
Asymptotic Giant Branch (AGB). The AGB phase consists of an early
period (Early AGB, E-AGB), in which the hydrogen-burning shell is
quiescent, and a subsequent, longer stage in which both shells are
active (Thermally Pulsing AGB, TP-AGB). The so-called AGB
phase-transition (Renzini 1981; Renzini \& Buzzoni 1986) corresponds
to the epoch at which a large amount of fuel is burned during the
TP-AGB phase. As a consequence, a sharp increase of the total
bolometric and infrared luminosity of the population is expected. The
AGB phase transition starts at ages $\sim 200\,$--$\,300$~Myr (see Iben \&
Renzini 1983; Renzini \& Voli 1981) lasts less than $\sim 1$~Gyr and
has sizable effects on the spectral energy distribution of the
population.

The inclusion of the TP-AGB contribution in population synthesis
studies is complicated by the poor knowledge of the real mechanisms
driving the evolution (see Girardi \& Bertelli 1998 for a
thorough investigation). Uncertainties in mass loss, mixing, and
efficiency of hydrogen burning at the bottom of the convective
envelope (known as the envelope burning process) together prevent pure
theory from predicting the amount of fuel (see Section 3.1) burned during
this phase (Renzini \& Buzzoni 1986).  Star clusters in the Magellanic
Clouds cover the interesting range in ages and offer the opportunity
to calibrate these effects.

Maraston (1998) presents models for Simple Stellar Populations (SSPs)
in which the TP-AGB contribution to the total energy as a function of
age is calibrated with the observed contribution in Magellanic Cloud
clusters. These SSP models simultaneously match the optical and
infrared colours of the intermediate age clusters. Unfortunately the
AGB phase transition in the Magellanic Cloud clusters is the only
observational counterpart of theoretical computations, allowing the
calibration of models with metallicity $\sim 0.5~Z_{\odot}$. The dependence
on $Z$ may be taken into account using the theoretical prescriptions
of Renzini \& Voli (1981).  This is explored in Maraston (1998) for
solar metallicity models.  For this work we compute intermediate age
SSPs models for various metallicities taking as a basis the
calibration by Maraston (1998) and adopting the prescriptions by
Renzini \& Voli (1981) to describe the influence of $Z$ on the AGB phase. 
The next section reports the adopted procedure in more detail.

\subsection{Modelling the different chemical compositions}

For the present work, new sets of SSP models with metallicities 
$0.5~Z_{\odot}$ ([Fe/H$]=-0.33$) and $2~Z_{\odot}$ ([Fe/H$]=0.35$), and ages
$t\ge 55$~Myr and $t\ge 100$~Myr respectively, are computed with the
evolutionary population synthesis code described in Maraston
(1998). The code is based on the fuel consumption theorem (Renzini \&
Buzzoni 1986) which allows a correct evaluation of the energetics of
the post Main Sequence phases.  The fuel consumption theorem states
that the contribution of any post Main Sequence phase $j$ to the total
luminosity, $L_{\rm j}^{\rm bol}(t,Z)$, is proportional to the amount
of fuel burned in that phase, $F_{\rm j}$, through the relation \\
$L_{\rm j}^{\rm bol}(t,Z) = 9.75\times 10^{10} b(t,Z)F_{\rm
j}(t,Z)$.\\ $b(t,Z)$, the ``evolutionary flux'', is the rate of
evolution off the Main Sequence, the fuel $F_{\rm j}$ is the amount of
hydrogen and /or helium (in $M_{\sun}$) burned during the phase $j$ by
the evolutionary mass corresponding to the age $t$ of the population,
which is the turn-off mass $M_{\rm TO}(t,Z)$.

The input stellar tracks from the dwarf Main Sequence up to the end of
the E-AGB phase, are from Bono et al. (1997) and S.~Cassisi (1999,
{\it private communication}).  Synthetic stellar spectra as functions
of effective temperature, gravity and metallicity, for $T_{\rm e}\ga
3500~{\rm K}$, are taken from the spectral library compiled by
Lejeune, Cuisinier and Buser (1998). Empirical colours for C-and
M-type AGB stars are taken from various sets of observational data in
the Magellanic Clouds, as described in Maraston (1998). A wider
discussion of the complete set of new SSP models is given in Maraston
(2000, in preparation). Here we concentrate on the intermediate-age
SSPs used in the present analysis.  The Initial Mass Function (IMF) of
these models, unless explicitly stated, is a power-law with Salpeter
exponent over the whole mass range ($0.1\,$--$\,100~M_{\sun}$), because the
impact of the exact exponent of the IMF on broadband colours is small,
particularly for young stellar systems (see Maraston 1998).

The fuel consumption during the TP-AGB phase as a function of
metallicity is computed on the basis of the calibrated TP-AGB fuel for
$Z=Z_{\odot}$ (Maraston 1998). The adopted procedure is as
follows. For a population of age $t$ and metallicity $Z$, the TP-AGB
fuel is a fraction of the envelope mass at the onset of the TP-AGB
phase, $M_{\rm env}^{\rm TP-AGB}(t,Z)$, that in turn depends on the
present core mass. Both quantities are extracted from the appropriate
evolutionary track. The calibrated fuel $F^{\rm TP-AGB}(t)$ for
$Z=Z_{\odot}$ given in Table~1 of Maraston (1998) allows the
computation of the fraction of fuel burned as a function of $M_{\rm
env}^{\rm TP-AGB}(t)$ for $Z=Z_{\odot}$. The metallicity dependence of
the envelope mass is known.  Interpolation in $M_{\rm env}^{\rm
TP-AGB}(t,Z)$ then allows the evaluation of the TP-AGB fuel for
various metallicities, $F^{\rm TP-AGB}(t,Z)$.

\begin{figure}
\resizebox{\hsize}{!}{\includegraphics{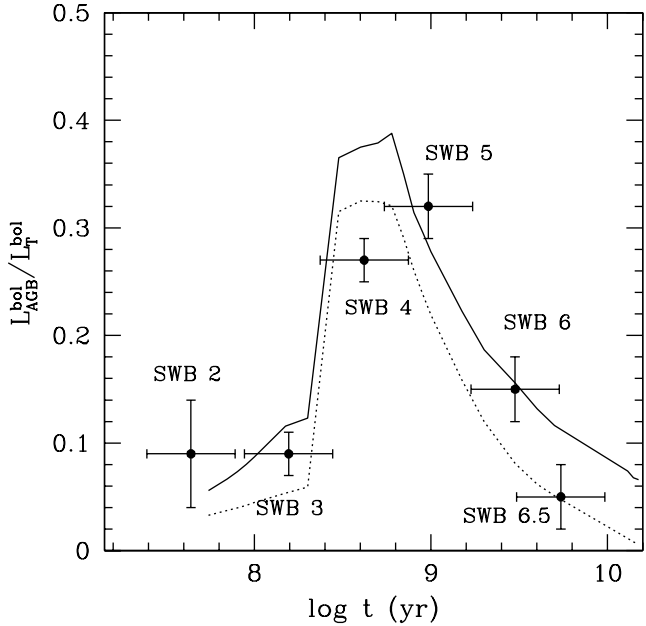}}
\caption{The synthetic AGB contribution to the total bolometric
luminosity compared with the LMC globular cluster data from Frogel,
Mould \& Blanco (1990). The contribution from the TP-AGB alone is
separately shown as a dotted line.}
\label{agbbol}
\end{figure}

The core mass, and therefore the envelope mass $M_{\rm env}^{\rm
TP-AGB}$ at the onset of the TP-AGB phase, do not strongly depend on
metallicity for any initial mass. This effect together with the
Magellanic cluster calibration, results in a total TP-AGB fuel which
is fairly independent of metallicity. Instead the metal content plays
a role in determining the spectral type of the stars burning the total
fuel, as will be explained later.  Fig.~\ref{agbbol} shows the
calibration of the synthetic AGB contribution to the total bolometric
luminosity as a function of age, evaluated with the fuel consumption
theorem.  The data points in Fig.~\ref{agbbol} are the observed AGB
contributions in Magellanic Cloud clusters (data from Frogel, Mould \&
Blanco 1990), grouped in age bins adopting the classification scheme
of Searle, Wilkinson \& Bagnuolo (1980, SWB). Within each bin, the
luminosities of all the clusters' AGB stars are added, and then
divided by the sum of all the integrated luminosities of the same
clusters. This procedure minimizes the large stochastic fluctuations
(due to the sparsely populated AGBs) between clusters having a same
SWB type (i.e. age). The AGB phase transition is observed among the
SWB groups $3\,$--$\,5$.  The figure is the analogue of Fig.~3 in
Maraston (1998) except that the data are now compared with models of
$Z = 0.5~Z_{\odot}$, more appropriate for Magellanic Cloud clusters.

The next step in computing realistic colours for SSPs is calibrating
the distribution of the total TP-AGB fuel among the various effective
temperatures. The TP-AGB phase is populated by stars of spectral type
C (carbon stars) and M, the production of C stars being a function of
both the initial stellar mass (hence of the SSP age) and the
metallicity (cf. Renzini \& Voli 1981).  From the data for Large
Magellanic Cloud clusters (Frogel, Mould \& Blanco 1990, Table~4),
Maraston (1998) evaluated the contributions of C and M stars to the
bolometric light for each age bin. The calibration, reported in
Table~2 and Fig. 4 of Maraston (1998), shows that C-stars
characterized intermediate-age SSPs ($0.2\,$--$\,2$~Gyr). Very old
stellar populations (e.g. Galactic globular clusters and the Galactic
Bulge) do not show any evidence for C-stars, in agreement with the
theoretical expectations. The very red infrared colours of the
intermediate age Magellanic Cloud clusters are due to the presence of
luminous ($M_{\rm bol}<- 4$) carbon stars, which are absent in the
youngest and oldest clusters and have no effect on visible colours
(Persson et al. 1983).

The initial metal content $Z$ of the stellar population plays a key
role in determining the production and the characteristics of carbon
stars (Renzini \& Voli 1981).  By decreasing $Z$, the abundance of
oxygen in the envelope is lower, thus a lower amount of carbon has to
be dredged-up in order to achieve {\rm C/O} $ > $ 1 and create carbon
stars.  Therefore, a metal poor stellar population is expected to
produce many more carbon stars than a metal rich one of the same age.
Increasing $Z$ from $\sim 0.5~Z_{\odot}$ (appropriate for LMC) to
$2~Z_{\odot}$ reduces the fraction of carbon stars by a factor of
$\sim 4$ (Renzini \& Voli 1981; see also Marigo 1999). This scaling
has been adopted to distribute the fuel among C- and M-stars, as a
function of metallicity.
\begin{figure}
\resizebox{\hsize}{!}{\includegraphics{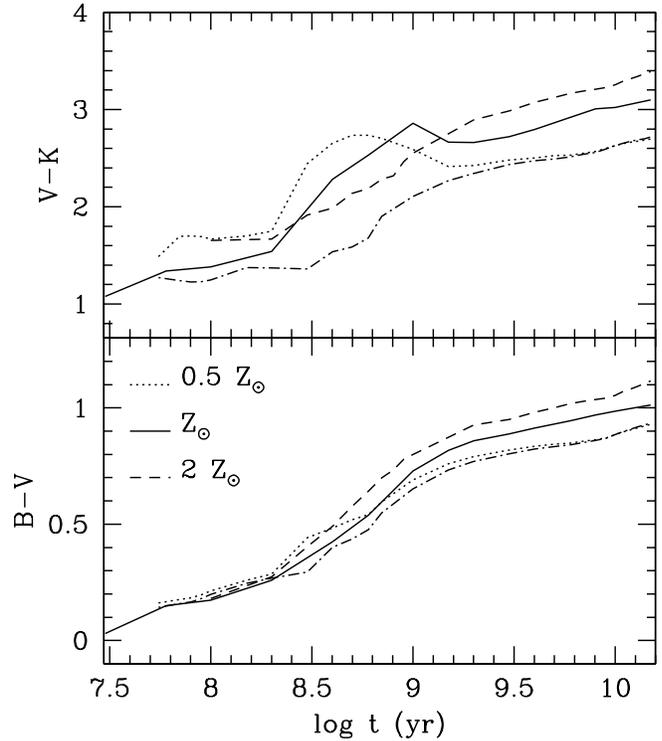}}
\caption{Time evolution of B$-$V (lower panel) and V$-$K (upper panel)
for SSPs of various metallicities (labelled in the lower panel). The
dot-dashed lines are models with $Z = 0.5~Z_{\odot}$ in which the
TP-AGB contribution is not taken into account.}
\label{colt}
\end{figure}
Fig.~\ref{colt} shows the resulting time evolution of the B$-$V and
V$-$K colours for SSPs with metallicity $0.5~Z_{\odot}$,~$Z_{\odot}$
and $2~Z_{\odot}$. Colours in other bands will be presented in
Maraston (2001, in preparation).

As already shown in Maraston (1998), the AGB phase transition has a
sizeable effect on SSP infrared colours. To give an idea of this
effect, an SSP (with $Z=0.5~Z_{\odot}$) is computed without including
the TP-AGB contribution (dot-dashed lines). This models also shows
that, as expected, the effect of AGB stars on optical colours is
negligible because these stars mainly radiate in the infrared.  Note
that the impact of the AGB phase transition {\it decreases} with the
increasing metallicity of the SSP in our modelling. This is because
less fuel is burned by C-like stars in a metal-rich population. In
Section 4.1 we show that models from other authors do not reproduce this
trend.

The SSP models displayed in Fig.~\ref{colt} are used to determine the
ages of the star clusters of NGC 7252, as shown in the next Section.

\section{The age distribution of NGC 7252 clusters}

\subsection{Evidence for the AGB phase-transition}

\begin{figure}
\centering
\psfig{figure=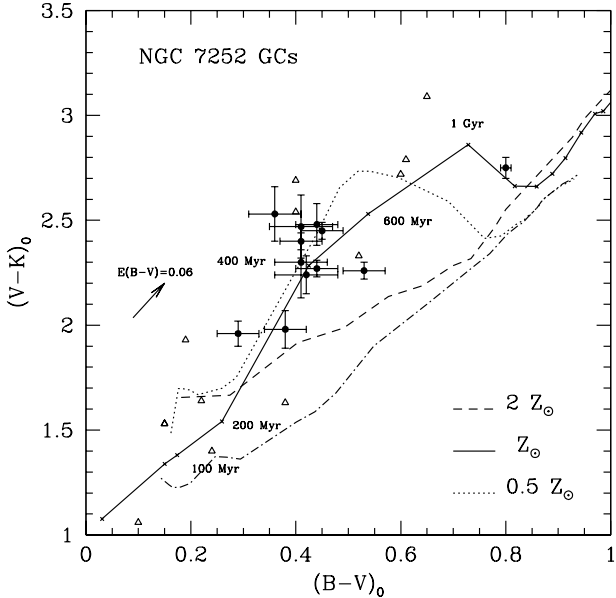,width=\linewidth}
\caption{The V$-$K vs. B$-$V diagram for the star clusters in NGC 7252
(filled circles). Also shown are LMC intermediate-age ($3$ \lapprox
SWB \lapprox $5.5$) GCs (open triangles), with B$-$V from van den Bergh (1981)
and V$-$K from Persson et al. (1983). SSP models for various
metallicities are over-plotted. The dot-dashed line is a model with
$Z=0.5~Z_{\odot}$ in which the TP-AGB contribution is not taken into 
account.}
\label{bvvk}
\end{figure}

The evolution of SSP optical colours, such as B$-$V, is monotonic with
age (see Fig.~\ref{colt}) because these mainly trace the evolution of
the turnoff region of the population. As will be shown in the next
section, the Balmer line strengths in contrast, have a maximum at
about $\sim 300$~Myr (cf. Fig.~\ref{fitline}). Therefore optical
colours help greatly in constraining the age of young SSPs.  For reasons
discussed in Section 3, an optical-to-infrared colour like V$-$K is
particularly useful for studying intermediate-age populations
dominated by AGB stars. Furthermore, V$-$K has the advantage of being
calibrated by the observed colours of LMC clusters, due to the many
sets of $K$-band observations that are available (see Maraston 1998 and
reference therein).

Fig.~\ref{bvvk} shows V$-$K vs. B$-$V for the star clusters in NGC
7252 (values from Table~\ref{kmags} and \ref{opticalmags}).  Also
shown are data for the intermediate-age GCs in Magellanic Clouds (open
triangles, references in the caption).  The arrow shows the total
Galactic foreground + internal extinction vector; we refer to Section
4.4 where the negligible impact of reddening on our age determination
is discussed.  The synthetic relations for the SSPs shown in
Fig.~\ref{colt} are over-plotted.  Some reference ages are indicated
for the $Z=Z_{\odot}$ model.

Fig. 5 clearly shows the AGB phase transition among the young star
clusters in NGC 7252. Both the optical and infrared colours of the
observed clusters are well described by our SSP models. The observed
range in infrared colours is not recovered if the TP-AGB contribution
is not taken into account in the synthesis (dot-dashed line model). In
the framework of the assumptions on the metallicity dependence of the
TP-AGB fuel (Section \ref{models}), half-solar or solar metallicities are
favoured for these GCs. In Section 5 a more insightful discussion on the
metallicities of selected clusters is given.
 
The young GCs of NGC 7252 lie in the narrow age range $200$~Myr to
 $500$~Myr, clustering around $\sim 300$~Myr. An important exception is
object W32 (B$-$V $\sim 0.8$~in Fig.~\ref{bvvk}), which seems to have
already completed the AGB phase transition.  The existence of such an
old cluster would impact on the assumed duration of the star formation process
triggered by merging (see Section 6).
\begin{figure}
\begin{center}
\begin{minipage}{0.3\textwidth}
\psfig{figure=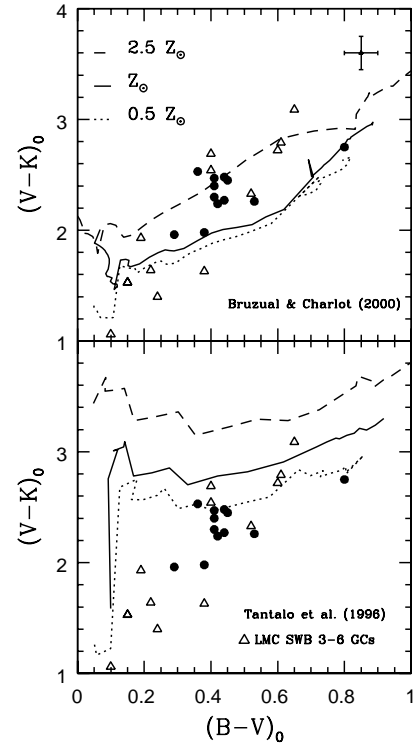,width=\linewidth}
\end{minipage}
\end{center}
\caption{The V$-$K vs. B$-$V diagram for the models of Bruzual \&
Charlot (2000, {\it private communication}, upper panel) and Tantalo
et al. (1996, lower panel). Data points are the same as in Fig.~\ref{bvvk}.}
\label{comp}
\end{figure}

As discussed in Section 3, the treat\-ment of the TP-AGB pha\-se in
po\-pu\-la\-tion syn\-the\-sis re\-pre\-sents a ma\-jor
dif\-fi\-cul\-ty, and intermediate-age SSPs models from various
authors show sizeable discrepancies (see Maraston 1998).
Fig.~\ref{comp} shows the V$-$K vs. B$-$V diagrams for the SSP models
of Bruzual \& Charlot (2000, {\it private communication}, upper
panel), and Tantalo et al. (1996, lower panel), with metallicities
indicated in the upper panel. The data are the same as in
Fig.~\ref{bvvk}. The trend of the synthetic V$-$K with $Z$, for
intermediate age SSPs is very different in the various models. The
Tantalo et al. (1996) models are redder at increasing metallicity and
do not show evidence for the AGB phase transition in V$-$K. As already
noted in Maraston (1998), the jump in V$-$K shown by the Tantalo et
al. (1996) models corresponds to very blue B$-$V ($\sim 0.15$~mag)
colours, and is in strong disagreement with the observations.  The
models by Bruzual and Charlot scale similarly with metallicity,
predicting redder colours with increasing $Z$.  However these models
are much bluer than those of Tantalo et al. (1996) and do not show an
appreciable difference between the solar and half-solar metallicity
cases.  We note that the infrared luminosity of intermediate age SSPs
in the most recent version of the Bruzual \& Charlot models is lower
than the in Bruzual \& Charlot (1996) models. This worsens the
comparison with the LMC data shown in Maraston (1998). In the Bruzual
\& Charlot models shown here, intermediate-age LMC GCs with B$-$V~$>
0.4$ are recovered by the $2.5~Z_{\odot}$ SSP, overestimating the real
metallicity of these clusters by a factor of $\sim$ 5. The GCs of NGC
7252 (filled dots) are of intermediate age, with $Z \simeq
2.5~Z_{\odot}$ according to BC2000 models, and are of intermediate
age, with $Z \la 0.5~Z_{\odot}$ according to the models by Tantalo et
al. (1996).

\subsection{Ages from photometry}
\label{photage}
\begin{table*}
\centering \begin{minipage}{130mm}
\caption[]{Photometric cluster ages (in Myr) determined on the
SSPs with metallicities indicated in the first line. The uncertainties
reflect the photometric errors. The last column contains the photometric age
derived by Schweizer \& Seitzer (1998) using V$-$I for $Z=Z_{\odot}$.}
\label{agephot}
\begin{tabular}{l c c c c c c c c l}
\hline
& \multicolumn{2}{c}{$0.5~Z_{\odot}$} &
\multicolumn{2}{c}{$Z_{\odot}$} & \multicolumn{2}{c}{$2~Z_{\odot}$} & 
$Z_{\odot}$ \\
Name & B$-$V & V$-$K & B$-$V & V$-$K & B$-$V & V$-$K & V$-$I (Schweizer \& Seitzer 1998) \\
\hline
W3 & 320 $\pm 70$ & 300 $\pm 10$ & 440 $\pm 60$ & 530 $\pm 30$ & 350
$\pm 50$ & 890 $\pm 30$ & 420 $\pm 110$ \\
W6 & 300 $\pm 60$ & 270 $\pm 10$ & 420 $\pm 60$ & 390 $\pm 20$ & 340
$\pm 50$ & 690 $\pm 70$ & 420 $\pm 90$ \\
W19  & -- & 280 $\pm 30$ & -- & 410 $\pm 100$ & -- & 750 $\pm 200$ &
-- \\
W22  & 300 $\pm 60$ & 300 $\pm 40$ & 420 $\pm 60$ & 550 
$\pm 90$ & 340 $\pm 40$ & 920 $\pm 110$ & -- \\
W24 & 280 $\pm 60$ & 300 $\pm 50$ & 380 $\pm 90$ & 540 $\pm 130$ &
310 $\pm 60$ & 910 $\pm 170$ & -- \\
W25  & 240 $\pm 30$ & 340 $\pm 60$ & 300 $\pm 60$ & 600 
$\pm 120$ & 260 $\pm 40$ & 970 $\pm 190$ & -- \\
W26  & 550 $\pm 120$ & 270 $\pm 10$ & 580 $\pm 70$ & 390 $\pm 20$ &
440 $\pm 50$ & 670 $\pm 60$ & 510 $\pm 70$ \\
W27 & 260 $\pm 30$ & 230 $\pm 10$ & 330 $\pm 60$ & 300 $\pm 30$ & 280
$\pm 40$ & 390 $\pm 80$ & -- \\ 
W30 & 280 $\pm 30$ & 290 $\pm 10$ & 380 $\pm 60$ & 480 $\pm 30$ & 310
$\pm 40$ & 850 $\pm 30$ & 400 $\pm 90$ \\ 
W31 & 200 $\pm 40$ & 230 $\pm 10$ & 230 $\pm 40$ & 300 $\pm 20$ & 210 $\pm
40$ & 360 $\pm 60$ & 230 $\pm 40$ \\
W32  & 2300 $\pm 300$ & $\sim 2000$ & 1400 $\pm 60$ & 1250 $\pm 130 $ &
1000 $\pm 50$ & 1460 $\pm 140$ & -- \\ 
S114 & 280 $\pm 70$ & 270 $\pm 10$ & 390 $\pm 90$ & 380 $\pm 40$ &
 320 $\pm 60$ & 650 $\pm 140$ & 370 $\pm 110$ \\
\hline
\end{tabular}
\end{minipage}
\end{table*}

The ages for the individual NGC 7252 clusters are obtained by
interpolating the model colours (and ages) shown in Fig.~\ref{colt} for
the observed B$-$V and V$-$K colours (Table~\ref{kmags} and
\ref{opticalmags}), separately.

The results are reported in Table~\ref{agephot}, for metallicities
$0.5$, $1$, $2~Z_{\odot}$. The uncertainties on the ages reflect the
photometric errors given in Table \ref{kmags} and \ref{opticalmags}.
If the values are preceded by the $\sim$ symbol, the observed colour
is larger (or smaller) than the maximum (minimum) model colour. In
this case, no extrapolations are made. The quoted age corresponds to
the model which is closest to the observed colour.

The last column gives the age derived by Schweizer \& Seitzer (1998)
from the {\em HST} V$-$I photometry of Miller et al. (1997), using the
1996 release of the Bruzual \& Charlot models with solar
metallicity. The ages from Schweizer \& Seitzer (1998) agree well with
ours, as determined from B$-$V (for solar metallicity).

\begin{figure}
\resizebox{\hsize}{!}{\includegraphics{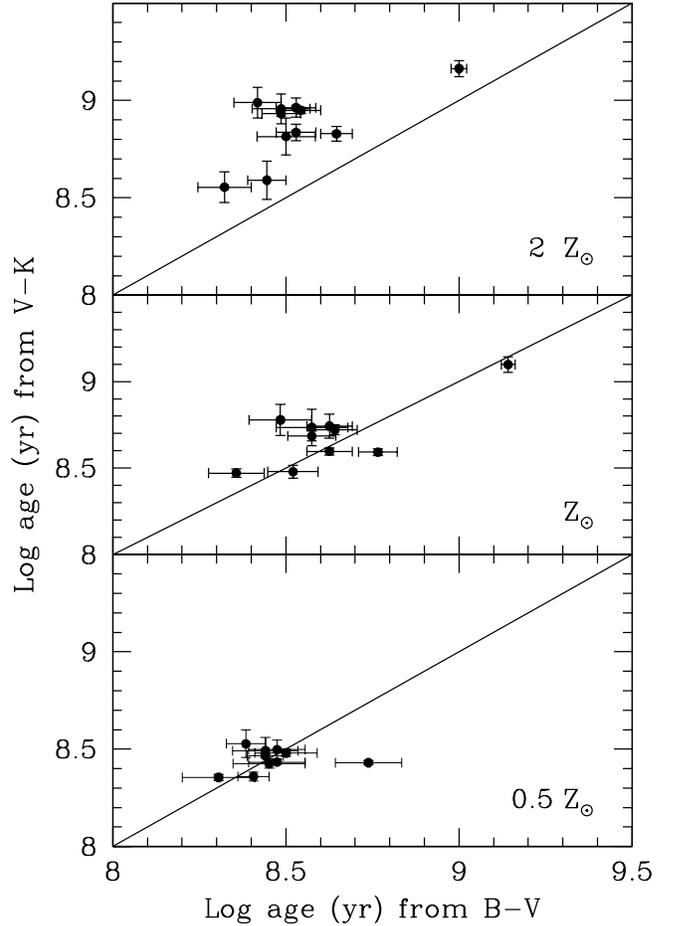}}
\caption{Comparison between the ages derived from B$-$V and V$-$K.}
\label{consage}
\end{figure}

The comparison between the ages from B$-$V and V$-$K given in
Table~\ref{agephot} is shown in Fig.~\ref{consage} for the various
metallicities. The models with $Z=0.5~Z_{\odot}$ provide the best
agreement between the ages derived from B$-$V and the ones derived from V$-$K. 
For some clusters, solar metallicity models (central panel)
also provide a good match (cf. Table~\ref{agephot}). Our models with a
higher metallicity ($Z=2~Z_{\odot}$, upper panel) fit the data
significantly less well. In fact, due to the lower amount of fuel
burned by carbon stars at increasing metallicity (see
Section \ref{models}), infrared colours are bluer for a given optical
colour and the ages derived from V$-$K are systematically larger than the
ones derived from B$-$V.  We note again that, due to the lack of
calibrations for the theoretical predictions, conclusions on the
metallicities of these clusters are only tentative.

As a final remark, it is important to note that the photometric ages
of Table \ref{agephot} do not depend on the particular choice of the
Salpeter IMF because, as anticipated in Section 3.1, broadband colours
are fairly insensitive to the IMF slope.  To quantitatively explore
the effect of the IMF, we consider an IMF biassed towards large masses
with exponent $-1.5$~(in the notation where Salpeter is $-2.35$) and
metallicity $Z=0.5~Z_{\odot}$.  The impact on broad-band colours and
mass-to-light ratios of an IMF of this kind is discussed in Maraston
(1998). Recalling the results, the optical and IR colours of a
giant-dominated SSP are {\it redder} than those of a Salpeter one for
ages $\la 4$~Gyr, the maximum difference being $\sim 0.04$~mag and
$\sim 0.12$~mag for B$-$V and V$-$K respectively.  For ages older than
4 Gyr, these colours become bluer than those for a Salpeter IMF by
$\sim 0.01$~mag and $\sim 0.06$~mag respectively.  The photometric
ages derived from the SSP with a giant-dominated IMF are lower by
$\sim 20\,$--$\,30$~Myr (e.g.~for W3 we obtain an age of $\sim 280$~Myr)
than the ones derived from the SSP with a Salpeter IMF. The effect of
the IMF is therefore negligible for the purposes of age determination.

\subsection{Photometric versus spectroscopic ages}
\label{compage}

It is interesting to check how the ages derived from photometry
compare to the ones spectroscopically determined in the framework of
the same modelling. With this aim the strengths of the Balmer lines
H$\beta$, H$\gamma$~ and H$\delta$ were measured on the spectra of
the SSP models described in Section 3.1.  The same procedure (spectral
resolution and bandpass definitions) was adopted as was used for the
observed spectra (Section 2.5, Table 3).

\begin{table*}
\centering
\caption[]{Spectroscopic cluster ages (in Myr).  Metallicities of the
models are given in the first line. Uncertainties are computed using
the errors given by Schweizer \& Seitzer (1998).}
\label{agespec}
\begin{tabular}{l c c c c c c c c c l}
\hline & \multicolumn{4}{c}{$0.5~Z_{\odot}$} &
\multicolumn{4}{c}{$Z_{\odot}$} \\ 
Name & H$\beta$ & H$\gamma$ & H$\delta$ & $t_{\rm av}$ & H$\beta$ & H$\gamma$ &
H$\delta$ & $t_{\rm av}$ \\ 
\hline 
W3 & $\sim$ 400 & 250 $\pm 10$ & $\sim$ 400 & 250 $\pm$ 10 & 
420 $\pm 20$ & 600 $\pm 10$ & $\sim$ 400 & 510 $\pm$ 10 \\ 

W6 & 310 $\pm 70$ & 200 $\pm 90$ & $\sim$ 400 & 260 $\pm$ 80 
& 600 $\pm 90$ & 640 $\pm 40$ & 430 $\pm 30$ & 560 $\pm$ 50 \\ 

W26 & 800 $\pm 600$ & 710 $\pm 200$ & $\sim$ 400 & 760 $\pm$ 400 
& 780 $\pm 480$ & 650 $\pm 170$ & $\sim$ 400 & 710 $\pm$ 320 \\ 

W30 & $\sim$ 400 & 270 $\pm 100 $ & $\sim$ 400 & 270 $\pm$ 100 
& $\sim$ 400 & 580 $\pm 140$ & $\sim$ 400 & 580 $\pm$ 140 \\ 

W31 & $\sim$ 60 & $\sim$ 60 & $\sim$ 400 & -- & $\sim$ 30 
& 30 $\pm$ 300 & $\sim$ 400 & 30 $\pm$ 300 \\ 

S114 & $\sim$ 60 & 180 $\pm 180$ & 220 $\pm 200$ & 200 $\pm$ 190 
& $\sim$ 30 & 110 $\pm$ 130 & 140 $\pm$ 1000 & 130 $\pm$ 560 \\ 
\hline
\end{tabular}
\\
\centering
\begin{tabular}{l c c c c c l}
\hline \multicolumn{4}{c}{$2~Z_{\odot}$} & Schweizer \& Seitzer (1998) 
($Z_{\odot}$) \\ 
 H$\beta$ & H$\gamma$ & H$\delta$ & $t_{\rm av}$ & all Balmer \\ 
\hline 
320 $\pm 30$ & 390 $\pm 10$ & $\sim$ 300 & 360 $\pm$ 20 
& 540 $\pm$ 30 \\ 

410 $\pm 50$ & 420 $\pm 30$ & 310 $\pm 40$ & 380 $\pm$ 40 & 580 $\pm$ 50 \\ 

540 $\pm 300$ & 430 $\pm 100$ & $\sim$ 300 & 490 $\pm$ 200 
& 530$^{+300}_{-200}$ \\ 

$\sim$ 300 & 380 $\pm 70$ & $\sim$ 300 & 380 $\pm$ 70 & 470 $\pm$ 40 \\ 

630 $\pm 1200$ & 510 $\pm 250$ & $\pm$ 500 & 570 $\pm$ 700 
& 530$^{+270}_{-160}$ \\ 

730 $\pm$ 170 & 440 $\pm$ 90 & 430 $\pm$ 190 & 530 $\pm$ 150 
& 1100 $\pm$ 300 \\ 

\hline
\end{tabular}
\end{table*}

\begin{figure*}
\psfig{figure=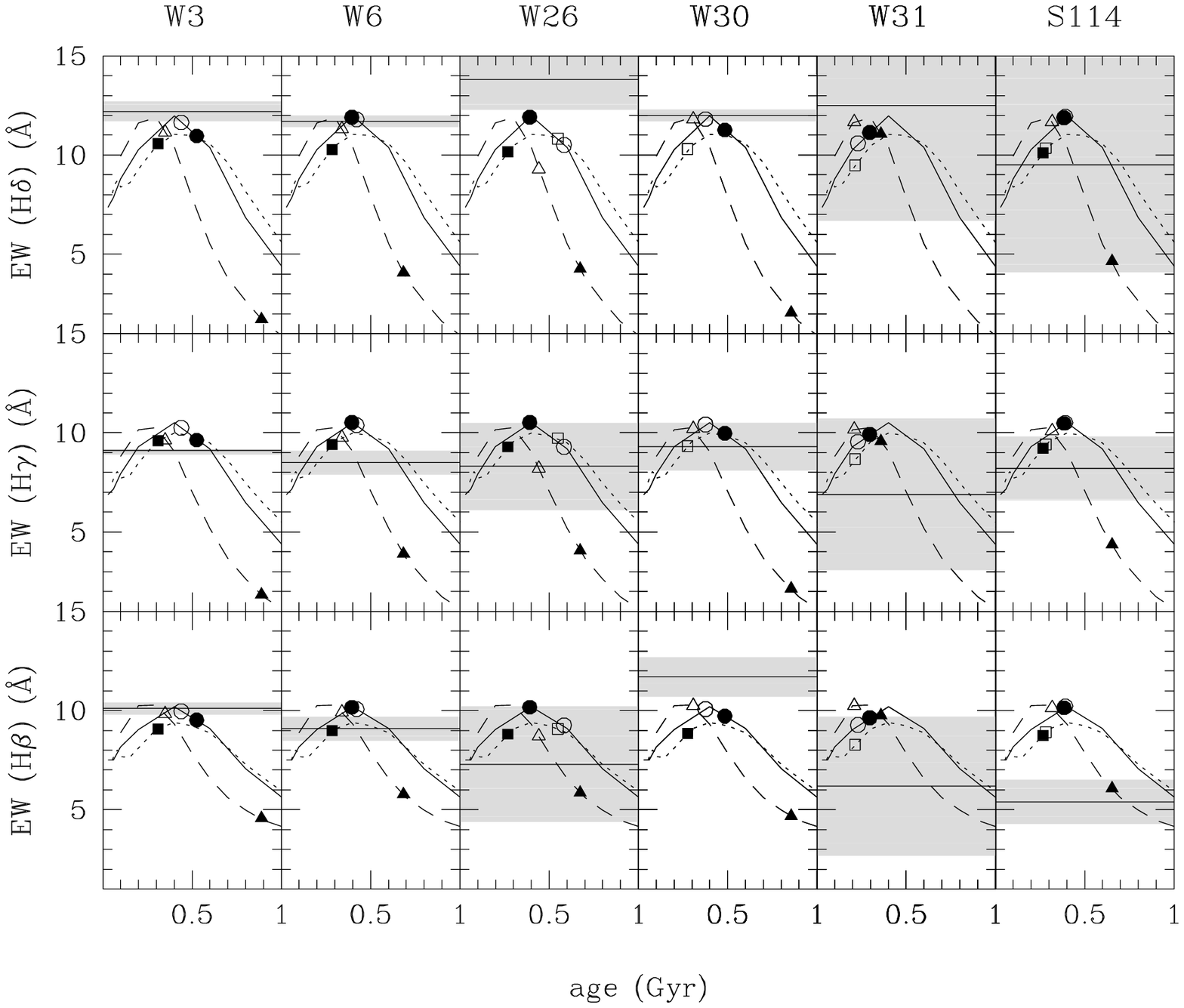}
\caption{Comparison between photometric and spectroscopic ages.  EWs
measured from the observed spectra (named on the top of each panel)
are represented as shaded areas, with widths of the same size as the
errors of Table 3. Curves are synthetic EWs measured from a set of
model spectra for SSPs with metallicity $0.5~Z_{\odot}$ (dotted line),
$Z_{\odot}$ (solid line) and $2~Z_{\odot}$ (dashed line). Symbols on
each curve show the photometric ages derived on the same models from
B$-$V and V$-$K (open and filled symbols, respectively, cf. Table 4). 
Squares are used for $0.5~Z_{\odot}$, circles for $Z_{\odot}$ and
triangles for $2~Z_{\odot}$.}
\label{fitline}
\end{figure*}

The results are given in Table~\ref{agespec}.  The ages are obtained
by interpolating the model EWs (and ages) for the various
metallicities (given in the first line) for the observed EWs.  In case
of two possible solutions for the age (cf.~Fig.~\ref{fitline}), the
age derived from the photometry is used to choose between the two
options. As for Table 4, if the values are preceded by the $\sim$
symbol, the observed EW is larger (or smaller) than the maximum
(minimum) model EW. In this case, the quoted age corresponds to the
model which is closest to the observed EWs.  In order to guide the
comparison, a value for the average Balmer age ($t_{\rm av}$) is also
given in Table 5.  This is simply obtained by averaging the values
derived from the three Balmer lines. The approximate values ($\sim$)
are not taken into account in the average.  The uncertainties are
computed using only the errors on the EWs given by Schweizer \&
Seitzer (1998) (see Table 3), and not taking into account any
systematic errors in the models.

The comparison between the spectroscopic and the photometric ages
(Table~\ref{agephot}) is shown in Fig.~\ref{fitline}.

Spectroscopic and photometric ages are consistent when the symbols lie
inside the shaded regions.  Note that the object S105, for which we do
not have photometry, is not included in the comparison of
Fig.\ref{fitline}.  In general, the average spectroscopic ages are in
good agreement with the ones derived from photometry, within the large
errors affecting most of the spectra.  Because of this, the photometry
greatly helps in constraining the age for poorly observed spectra. An
example is object S114, for which Schweizer \& Seitzer (1998) could
not assess the age.

\subsection{Internal reddening}

Our infrared photometry (Table 1) and the optical photometry from the
literature (Table 2) are corrected for a Galactic foreground
extinction of $E(B-V)=0.012$ (see Section 2.2 and 2.3). In this Section
we explore the impact of internal reddening on the cluster ages
derived from photometry.

NGC 7252 shows evidence for dust in the central region ($r\la 2.5$
kpc), associated with spiral-arm like structures (Whitmore et
al. 1993, Miller et al. 1997). As discussed in these studies, most of
the clusters belonging to the ``outer'' galaxy region (6\arcsec\ to
$\sim$ 120\arcsec ), are not found in dusty regions. Therefore it is
likely that our magnitudes are not severely affected by internal
reddening. Moreover, the IUE spectrum of NGC 7252 is rather blue,
which is also indicative that the actual internal reddening is likely
to be not too large (D. Calzetti, {\it private communication}).  From
the Third Reference Catalog of Bright Galaxies (RC3, de Vaucouleurs et
al. 1991) the {\it global} $E(B-V)$ is $\sim 0.06$. This estimate
contains the differential Galactic foreground extinction as determined
by Burstein \& Heiles (1984), and the differential internal
extinction, which is evaluated according to the galaxy morphological
type (see RC3).

To explore the effect of the internal reddening, we additionally
correct our optical and infrared magnitudes by $E(B-V)=0.06$ and
$E(V-K)=0.16$. This reddening vector (see Fig.~\ref{bvvk}) is nearly
parallel to the models, implying that the derived ages decrease, but
without dramatically changing the metallicity of the best fitting
model (see Section 4.2).  We find that the photometric ages are $\sim$ 10
\% lower using V$-$K and 14 \% lower using B$-$V, for the models with
$Z=0.5~Z_{\odot}$. This variation represents an upper limit, as the
actual internal reddening is likely to be lower than our conservative
assumption. Hence we conclude that the optical-to-infrared colour
distribution of our cluster sample mainly reflects the present stellar
population.

\subsection{Stochastical fluctuations in the near-IR}

The bright infrared magnitudes of intermediate-age populations
dominated by the TP-AGB phase are potentially affected by
stochastic fluctuations due to the small duration of this phase
($t\sim10^6{\rm yr}$). This in turn implies small numbers of AGB
stars. These effects are seen among the Magellanic Clouds clusters
with a similar SWB parameter (i.e. a similar age) as a spread in the
V$-$K colours. The two-colour diagram of Fig.~\ref{bvvk} already
proves that large stochastical fluctuations are not present in our
sample, which has a rather small scatter. This is reinforced by the
close agreement between the ages determined using B$-$V and V$-$K,
shown in Fig. 7. The small fluctuations are a consequence of the high
luminosity of most of our clusters, which implies a large sampled
number of AGB members.  It is then interesting to estimate how many
AGB stars are likely present in our clusters. Their number can be
evaluated using the fuel consumption theorem.

In a stellar population of given age the number $N_{\rm j}$ of stars
present in a certain evolutionary phase j is proportional to the total
bolometric luminosity of the population ($L_{\rm T}$) and to the
duration of the phase ($t_{\rm j}$), i.e.  \\ $N_{\rm j}=B(t)L_{\rm
T}t_{\rm j}$.\\ The coefficient of proportionality $B(t)$ is the {\it
specific evolutionary flux} of the population (Renzini 1981; Renzini
\& Fusi Pecci 1988; Maraston 1998), which is the number of stars
entering or leaving any post-Main Sequence evolutionary stage per year
and per solar luminosity of the population. Following Maraston (1998,
see also Renzini 1999), the bolometric luminosity can be replaced by
the total luminosity in any band $ L_{\lambda}$, using the {\it
bolometric correction factor} appropriate to the age and the
metallicity of the population. The bolometric correction factors are
by-products of population synthesis and are defined as $L_{\rm
T}/L_{\lambda}$.  Using the value $t\sim10^6$ for the TP-AGB lifetime
and the photometric ages (for $Z=0.5~Z_{\odot}$) given in Table 4,
we evaluate the number of TP-AGB stars for every cluster with a given
$V$ and $K$ magnitude. The observed $V$ and $K$ magnitudes are
converted to luminosities adopting a distance modulus of 34.04
(assuming $H_{0}=75$ km s$^{-1}$
Mpc $^{-1}$, following Schweizer \& Seitzer 1998) and a Galactic foreground
extinction $A_{\rm K}$=0.36*$E(B-V)$ (see Section 2.2). The $V$ and $K$
magnitudes of the Sun are 4.83 and 3.41 respectively (Allen 1991).
\begin{table}
\centering
\caption[]{The expected numbers of TP-AGB stars for the clusters
listed in Table 1. The ages for each cluster are from Table 4 for the
model with $Z=0.5~Z_{\odot}$. Also quoted are the corresponding 
fluctuations.}
\label{stoflu}
\begin{tabular}{l c c c c c l}
\hline
Name & $N_{\rm j}(V)$ & $N_{\rm j}(K)$ & $1/{N_{\rm j}(V)}^{1/2}$ &
$1/{N_{\rm j}(K)}^{1/2}$ \\ 
\hline 
W3  &   4900 & 4700 & 0.01 & 0.01 \\
W6  &    950 &  841 & 0.03 & 0.03 \\
W19 &     73 &   73 & 0.12 & 0.12 \\
W22 &    153 &  153 & 0.08 & 0.08 \\
W24 &     56 &   56 & 0.13 & 0.13 \\
W25 &     70 &   72 & 0.12 & 0.12 \\
W26 &    490 &  300 & 0.05 & 0.06 \\
W27 &    100 &  100 & 0.10 & 0.10 \\
W30 &   1150 & 1200 & 0.03 & 0.03 \\
W31 &    290 &  315 & 0.06 & 0.06 \\
W32 &    220 &  290 & 0.07 & 0.06 \\
S114 &   225 &  222 & 0.07 & 0.07 \\ 
\hline
\end{tabular}
\end{table}
The expected numbers of TP-AGB stars are given in Table \ref{stoflu}.
The TP-AGB phase in our clusters is well populated, the faintest
clusters (W24 and and W19) having the smallest numbers of TP-AGB
stars. The expected stochastic fluctuations are given in the last
two columns and range between 1 \% and 13 \%. The corresponding magnitude
fluctuations are of the same order, therefore smaller or comparable to
the observational errors (see Table 1). A remarkable consistency is
evident between the star counts obtained using the $V$ and the $K$
luminosities. This is not surprising because our SSPs are able to
simultaneously reproduce the optical and IR colours very well, for
$Z=0.5~Z_{\odot}$. Note that the expected numbers of stars are not
significantly affected by changing the metallicity, because the
specific evolutionary flux remains almost unchanged (see Renzini 1998;
Maraston 2000) and the main driver is the large luminosities of the
clusters.

\section{Metallicities and abundance ratios}

\subsection{The young star clusters}

The direct determination of cluster metallicities by measuring the
strength of metallic lines (e.g. Mg, Fe, etc.) is complicated by two
effects.  The artificial broadening of the observed spectra, required in order 
to compare with the model ones, undesirably dilutes the metallic lines
which are typically weak in the spectra of young stellar
populations. This is aggravated by the low signal-to-noise of most of
the spectra (see Table~3 and discussion in Schweizer \& Seitzer 1998).
For these reasons we restrict the measurement of metallic lines to the
best observed object, W3. We extend the exercise to objects W6 and W30,
cautioning that these spectra are of much lower quality.

The equivalent widths of Mgb and Fe5270 are measured on the
observed-broadened and model spectra. The comparison is made by
measuring the indices on observed and model spectra exactly in the
same way (see Sec. 2.5). For convenience, we adopt the definitions of
line and continuum bandpasses given in Worthey et al. (1994). It is
important to note that these model indices are not computed by means
of the fitting functions given in Worthey et
al. (1994).

\begin{figure}
\resizebox{\hsize}{!}{\includegraphics{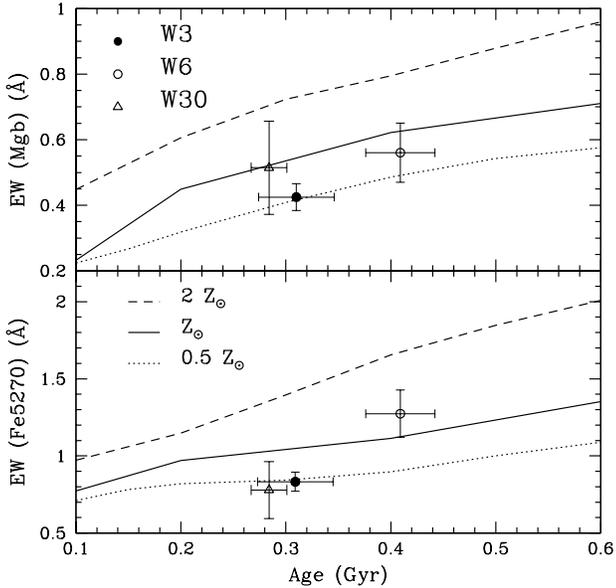}}
\caption{Fe5270 and Mgb EWs (\AA) for objects W3, W6 and W30, plotted
vs. the photometric ages for $Z=0.5,~1,~0.5~Z_{\sun}$ respectively.
Lines show the synthetic EWs for SSPs with various metallicities.}
\label{metal}
\end{figure}

Fig.~\ref{metal} shows the synthetic EWs of Fe5270 (bottom panel) and
Mgb (top panel) for SSPs with various metallicities, as functions of
age. The measured values for objects W3, W6 and W30 (symbols) are
plotted versus the photometric ages given in Table 4 (see below). The
vertical error bars are derived from the flux calibrated spectra by
scaling our measurements errors to our measured H$\beta$ errors, which in
turn are matched to the H$\beta$ errors given in Schweizer \& Seitzer
(1998) (see Table 3).

We now comment on the individual clusters.

An excellent agreement is found for object W3 between the measured EWs
(Fe5270 $\sim 0.83$~\AA, Mgb $\sim 0.43$~\AA) and the SSP with $Z=
0.5~Z_{\odot}$, using the photometric age derived from this same model
($\sim 300$ Myr). By contrast, the photometric ages inferred from the
$Z_{\odot}$ and the $2~Z_{\odot}$ models correspond to EWs larger than
the observed ones for both lines.  The consistency between the results
from photometry and spectroscopy indicates that W3 is a $\sim 300$~Myr
old, $0.5~Z_{\odot}$ metallicity object with high likelihood. Note
that measuring the Mgb and Fe EWs on the original, higher resolution
spectrum gives Fe5270 $\sim 0.84$~\AA\ and Mgb $\sim 0.48$~\AA\
respectively, consistent with the values measured on the broadened
spectrum.

In the case of object W6 (Fe5270 $\sim 1.27$~\AA\ and Mgb $\sim
0.56$~\AA), the best agreement is found when the average photometric age
derived from the $Z_{\odot}$ model is adopted.  The other two
metallicities agree less well. 

Finally, we obtain Fe5270 $\sim 0.78$~\AA\ and Mgb $\sim 0.51$~\AA\
for object W30. These values do not allow us to clearly discriminate
between half-solar and solar metallicity for this cluster (the former
being slightly preferred), but metallicities much above solar appear
to be ruled out.

Object W6 appears to be slightly more metal-rich than object W3. This
is in qualitative agreement with the findings of Schweitzer \& Seitzer
(1998), although they derive larger metallicities for both clusters.  

This may be due to the use by Schweitzer \& Seitzer (1998) of model
Lick indices based on the analytical fitting functions given in
Worthey et al. (1994). As also noted by these authors, these functions
are designed for old stellar populations, and it is not clear how well
these can be applied to younger stellar populations.

The abundance ratios of $\alpha$-elements to iron-group elements carry
information on the timescales over which star formation occurs,
because $\alpha$-elements (e.g., Mg, O, etc.) are produced in Type II
supernovae (short-lived progenitors), while a significant fraction of
iron comes from Type Ia supernovae (long-lived progenitors). A solar
Mg/Fe implies star formation timescales of the order 10~Gyr, while
[Mg/Fe]$\sim 0.2$ dex requires formation timescales $\sim 1$~Gyr
(e.g., Thomas, Greggio, \& Bender 1999). If the observed Mg
line-strength is stronger than predicted by SSP models based on solar
abundance ratios, an enhancement of $\alpha$-elements is likely to
exist, as is the case for luminous elliptical galaxies (see
Section 6.3).
 
 From Fig.~\ref{metal} it appears that object W3 is not overabundant
in $\alpha$-elements. This is what it is expected if the clusters form
out of the highly Fe-enriched interstellar gas of the progenitor
spirals (Thomas, Greggio, \& Bender 1999). This result is in agreement
with the prediction by Fritze-v Alvensleben \& Gerhard (1994a), that
any star clusters formed during mergers should have abundance ratios
typical of the interstellar medium in the progenitor spirals.

\subsection{The diffuse light of NGC 7252}

If the young globular clusters form out of the interstellar medium
associated with the merger event, it is interesting to investigate how
the overall galaxy spectrum compares with the ones of the newly formed
clusters.  The spectrum of the diffuse light is produced by a mixture
of stellar populations of various ages and metallicities, the light
being dominated by the last episode of star formation. The
interpretation by means of SSP models gives, by definition, only a
measure of the average population mix. This analysis is nonetheless
useful in order to identify the possible presence of features
strikingly different from the cluster spectra.
\begin{figure}
\resizebox{\hsize}{!}{\includegraphics{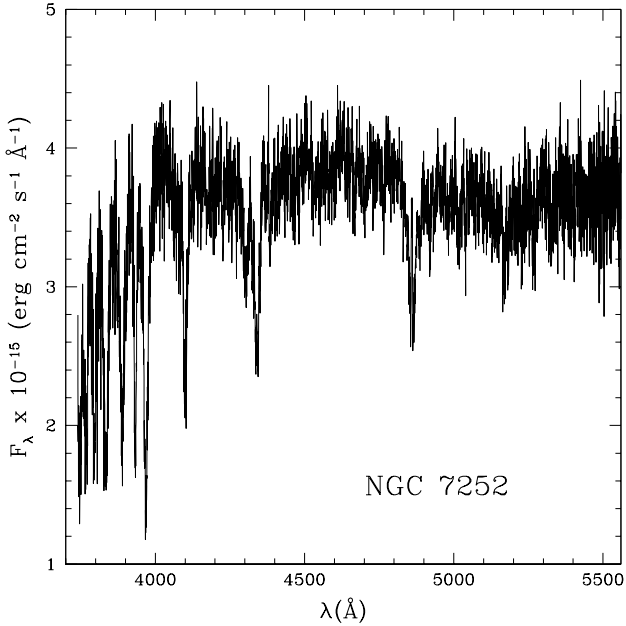}}
\caption{Nuclear (2''x2'') spectrum of NGC 7252 (F. Schweizer, {\it
private communication})}
\label{spegal}
\end{figure}
F. Schweizer kindly provided us with the spectrum of the central
region of NGC 7252 (Fig.~\ref{spegal}). The spectrum was taken in 1982
with the Las Campanas [Du Pont] 2.5-meter telescope, equipped with
Shectman's Reticon. The aperture was 2"x 2", centered on the nucleus,
and the integration was 2000 sec, with a high dispersion (1200 l/mm)
grating. This spectrum has been broadened (see Section 2.5) and the
Balmer lines plus some metallic lines have been measured, using the
same bandpass definitions as for the cluster spectra. For the
additional metallic lines Fe5335 and Mg$_2$ used here, the index
definitions given in Worthey et al. (1994) are adopted.  In the case
of the NGC 7252 nuclear spectrum, Mg$_2$ is preferred to Mgb
because the continuum of the latter can be increased by continuum
hydrogen emission, common in young regions. This would produce an
artificially deep Mgb line, because the continuum of the Mgb feature
is defined to be adjacent to the line itself. Mg$_2$ should suffer
less from this effect because its continuum bandpasses are far enough
from the absorption line.

The following values are obtained for the nuclear spectrum: H$\delta$
$\sim 7.52 \pm 0.3$~\AA, H$\gamma$ $\sim 6.62 \pm 0.6$~\AA, 
H$\beta$ $\sim 6.2 \pm 0.6$~\AA, Mg$_2$ $\sim 2.77 \pm 0.31$~\AA, 
Fe5270 $\sim 1.26 \pm 0.21$~\AA~and Fe5335~$\sim 0.96 \pm 0.19$~\AA.

\begin{figure}
\resizebox{\hsize}{!}{\includegraphics{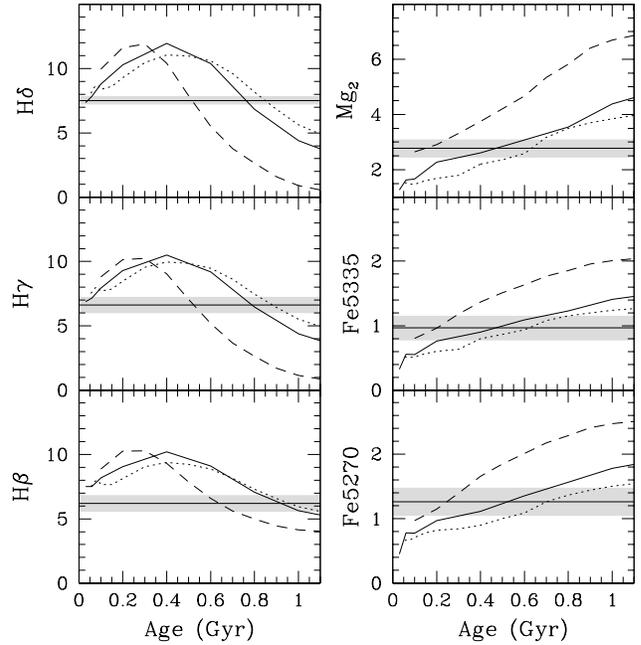}}
\caption{Time evolution of Balmer line (left-hand plots) and metallic
line (right-hand plots) EWs (\AA) for SSPs with metallicities 
$0.5~Z_{\odot}$ (dotted line), $Z_{\odot}$ (solid line) and $2~Z_{\odot}$
(dashed line). The values measured on the spectrum of the central
region of NGC 7252 (Fig.\ref{spegal}) are shown as straight lines.}
\label{linegal}
\end{figure}
In Fig.~\ref{linegal} these values (marked as straight lines) are
compared to the predictions of SSPs with various metallicities,
plotted as functions of age. Shaded areas represent errors.

The strengths of $H\gamma$ and $H\delta$ (left-hand panels)
consistently suggest an age of: $\sim 870$~Myr for metallicity
$0.5~Z_{\odot}$; $\sim 780$~Myr for solar metallicity; $t\sim 520$~Myr
for metallicity $2~Z_{\odot}$.  The slightly larger ages inferred from
$H\beta$ are because its value is lowered by filling from an emission
line (clearly visible in Fig.~\ref{spegal}).  The strengths of the
metallic lines (right-hand panels) also consistently constrain the
age, which is, however, systematically lower than the one obtained
from the Balmer lines, for every metallicity.  The following value are
found: $\sim 630$~Myr for metallicity $0.5~Z_{\odot}$; $\sim 480$~Myr
for solar metallicity; $t\sim 240$~Myr for metallicity $2~Z_{\odot}$. 
The smallest discrepancy amounts to 28 \% for the $0.5~Z_{\odot}$
model; within the error bars the ages from Balmer lines and from
metallic lines are consistent for the $0.5~Z_{\odot}$ model.

On average, the spectral features of the galaxy central diffuse light
are consistent with a rather young average population with metallicity
slightly sub-solar, in agreement with what is derived from the analysis
of the star clusters. As in the case of the star clusters, the
spectrum of the diffuse light does not show an enhancement in
$\alpha$-elements. Simulations of the chemical evolution of composite
stellar populations where a young component is born on top of an underlying
old population during a spiral--spiral merger (Thomas, Greggio \& Bender 1999) 
are in agreement with this result, in the sense that no
$\alpha$-enhancement is observed in the resulting composite population.

Note that our results disfavour an IMF substantially flatter than
Salpeter for the stars formed during the merger. A flat IMF would
imply a larger number of SN II and fast enrichment of the gas in
$\alpha$-elements. Therefore we would expect the newly formed stellar
populations to be $\alpha$-enhanced (Thomas, Greggio \& Bender 1999).

\section{Discussion}

\subsection{The recent star formation history of NGC 7252}

The derived ages and the age spread are compared with different models
in order to constrain the nature of the progenitors and derive the star
formation history during the merger event.

\subsubsection{The nature of the progenitors}

\begin{figure}
\resizebox{\hsize}{!}{\includegraphics{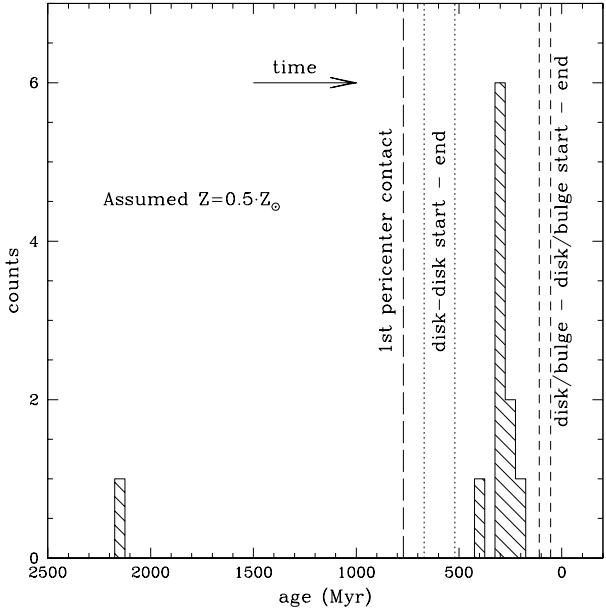}}
\caption{Histogram of the mean cluster ages (as derived from photometry,
assuming $Z= 0.5~Z_\odot$). Time runs from left to right, the
first pericenter contact $770$ Myr ago is marked as a dashed line. The
predictions from the simulations of Mihos \& Hernquist (1996) are shown
as dotted lines (marking the start and end of the main starburst in a
disk - disk merger) and as short dashed lines (start and end of the main
starburst in a disk/bulge - disk/bulge merger).}
\label{agehisto}
\end{figure}

The ages derived for the observed young clusters confirm the small
spread in age reported by Schweizer \& Seitzer (1998), see
Fig.\ref{agehisto}. Using different SSP models, we infer slightly
lower cluster ages, ranging between 300 and 500 Myr. We
confirm that cluster S114 for which Schweizer \& Seitzer could not
conclude an age, does indeed fall in the same age range as the other
clusters ( $300$~ -- $400$~Myr old, depending on the
metallicity).

As previously pointed out by Schweizer \& Seitzer (1998), the ages are
in reasonable agreement with predictions from N-body simulations for
NGC 7252 for the star formation during a disk-disk merger (Mihos \&
Hernquist 1996) which predict (for $H_0=75$ km s$^{-1}$~Mpc $^{-1}$)
that the first pericenter contact happened $\sim 770$~Myr ago, the
major burst of star formation occurring $\sim 100$~Myr years after
this encounter, lasting $\sim 150$~Myr. We note that the predictions
for star formation during the merger of bulge dominated disks differ
from the predictions for a disk -- disk encounter (Mihos \& Hernquist
1996). In bulge dominated mergers, galaxies appear stabilized against
bar modes which tend to delay the gas inflow: star formation is
predicted to happen $\sim 500$~Myr after the close encounter and to be
more violent but of shorter duration.  In Fig.~\ref{agehisto}
(cf.~Mihos \& Hernquist's Fig.~5) we show a histogram of the clusters
ages (as derived from photometry for $Z = 0.5~Z_\odot$) together
with the predictions for a disk--disk merger and a bulge dominated
merger. The beginning of the star formation does match either case: it
starts slightly later as expected for a disk-disk merger, but slightly
earlier than expected in a bulge dominated merger (although the
timescale of the simulation can be rescaled by the distance to NGC
7252 and the mass of the progenitors). The duration of the star
formation (taken as the dispersion of our ages $<$ 1 Gyr) is around 
$50$~Myr, closer to the prediction for a bulge dominated merger.  With
respect to the simulations, we cannot strongly conclude whether the
progenitors of NGC 7252 were bulge dominated or not, but see evidence
that they were not pure disks.

Fritze-von Alvensleben \& Gerhard (1994b) studied the integrated
properties of the stellar light and reached the conclusion that NGC
7252 is best modelled by the merger of two massive Sc type
spirals. Further, their study suggests that star formation took place
1.3 to 2 Gyr ago, well before the time of the first pericenter
contact, namely $0.6\,$--$\,0.7$~Gyr (Borne \& Richstone 1991). These
conclusions are in agreement with our observations.

\subsubsection{Are there clusters in the 1 to 2 Gyr range?}

One cluster (W32) stands out in colour as older than the others.
Unfortunately, no spectroscopy is available for it.

At face value, from the combination of its optical and near-infrared
colours and our plotted SSPs, W32 appears to be a cluster with an age
in the range 1 to 2 Gyr rather than being an old ($t\sim 10$~Gyr)
cluster belonging to the progenitors spirals. However, its optical
colours (B$-$V$=0.80\pm0.04$, V$-$I$=1.08\pm0.04$) could, according to
SSP models, still be compatible with a metal-poor, old (12 to 15 Gyr)
cluster.  For example, Maraston (1998) models for Z$=0.006$, 15 Gyr,
Gould et al.~(1997) IMF predict B$-$V$=0.87$, V$-$I$=1.08$. In this
case, the corresponding V$-$K should be $2.59$, instead of
$2.75\pm0.05$. Adding the near infrared colour therefore seems to
exclude an old ($10\,$--$\,15$~Gyr) age.  If W32 had an age of 15 Gyr its
total luminosity, for a distance modulus of $(m-M)=34.04$ (assuming
$H_0=75$ km s$^{-1}$ Mpc $^{-1}$), would be $M_V=-12.5$ mag which is
one to two magnitudes brighter than the brightest globular clusters in
the Milky Way and M31. There is no obvious reason for an abnormal
M/L$_V$ which could translate the high luminosity into a normal mass
(see Section 6.2).  We conclude that W32 is likely to be a cluster in
the age range 1 to 2 Gyr, and must have formed before the first
pericenter contact of the two merging galaxies.

Finally, we checked whether W32 is an exception by verifying if our
observed sub-sample is biased with respect to the total sample of
Miller et al.~(1997). A Kolmogorov-Smirnov test returns a probability
of $\sim$ 90\% that both samples are drawn from the same population.
Miller et al.'s is assumed to be unbiased with respect to object
colour in the range of interest, so that one could expect a few more
clusters with similar ages to W32.

\subsection{The cluster masses}

Interestingly the young clusters, as already discussed by Schweizer \&
Seitzer (1998), appear too bright by one to two magnitudes, compared to
massive globular clusters, for the assumed distance. 
Could the distance to NGC 7252 have been overestimated? 
The systemic velocity of $4828\pm3$ km s$^{-1}$
for NGC 7252 was measured by Schweizer (1982). This value is in
good agreement with the mean heliocentric velocity of 5083 km s$^{-1}$
and velocity dispersion of about 430 km s$^{-1}$ of the NGC7284/85 group, 
to which NGC 7252 belongs. Using the latest flow field model by the Key
project (Mould et al.~2000), this translates into a distance around 68
Mpc (assuming $H_0=75$ km s$^{-1}$ Mpc$^{-1}$), or $(m-M)\simeq 34.16$.
It is therefore excluded that the distance to NGC 
7252 has been significantly overestimated. Note that assuming $H_0=50 $ km
s$^{-1}$ Mpc$^{-1}$ would aggravate the problem.

Thus the young clusters in NGC 7252 {\it are} exceptionally luminous.  The
three brightest objects W3, W6 and W30 have $M_{\rm V}=-16.2, -14.4,
-14.6$, respectively (for $H_{0}=75$ km s$^{-1}$ Mpc $^{-1}$ and a
distance modulus $m-M$=34.04, Schweizer \& Seitzer 1998).  It is
interesting to evaluate the cluster masses $M$=$(M/L)\times L$,
adopting a model mass-to-light ratio. In order to compare with
present day, $\sim$ 15 Gyr old globular clusters, the fading of the
young clusters due to aging has to be considered. We concentrate on
object W3, which is the most luminous cluster.

Our synthetic stellar mass-to-light ratios take into account the
contribution from stellar remnants, as described by Maraston (1998,
see also Maraston 1999).  The photometric age for object W3 is $t\sim
300$~Myr (for $Z=0.5~Z_{\odot}$, see Table 4).  The corresponding
mass-to-light ratio in $V$ for a Salpeter IMF is $\sim 0.28$, which
translates in a present mass for W3 $\sim 7.2 \times
10^{7}~M_{\odot}$, a factor of $\sim 2.5$ lower than the value
estimated by Schweizer \& Seitzer (1998) ($1.8\times
10^{8}~M_{\odot}$).  The difference most likely results from Schweizer
\& Seitzer (1998) assigning an age of $\sim 500$~Myr and a solar
metallicity to W3.  Both the older age and the larger metallicity
increases the $M/L$ and hence the estimated mass. Part of the
difference comes from the use of a model $M/L$ (Bruzual \& Charlot
1996 in Schweizer \& Seitzer 1998), in which stellar mass loss is not
taken into account.

To explore the extent to which the mass can be lowered by changing the
IMF, we additionally consider two cases.  The first is a
giant-dominated IMF with exponent $-1.5$ (in the notation in which
Salpeter is $-2.35$). As already mentioned in Section 4.2, the age for W3
changes to $280$~Myr, corresponding to $M/L_{\rm V} \sim 0.19$,
hence a mass $4.9\times 10^{7}~M_{\odot}$. This is lower by $\sim$ 32
\% than the one derived for the Salpeter case.

The second option is an IMF which is flattened at the low-mass end. We
consider the one suggested by Gould, Bachall \& Flynn (1997), based on
{\em HST} observations of disk $M$ dwarfs. The shape of this IMF is $-0.9$ for
$ M \leq 0.6~M_{\odot}$, $-2.21$ for $0.6 < M/M_{\odot}\leq 1$ and $-2.35$
for $m/M_{\odot}\geq 1$ (in the notation in which Salpeter is $-2.35$).
The derived age for W3 is $\sim 300$~Myr, as in the Salpeter case, because
the effect on colours is negligible.  The corresponding mass-to-light
is $M/L_{\rm V} \sim 0.155$, hence the present mass for W3 is $4\times
10^{7}~M_{\odot}$. This is lower than the mass for the model
with a Salpeter IMF by 45 \%.

At an age of 15 Gyr, W3 will have a mass $\sim 6.3 \times
10^{7}~M_{\odot}$ for the Salpeter case ($M/L_{\rm V} \sim 6.8$),
$\sim 3 \times 10^{7}~M_{\odot}$ for the Gould et al. case ($M/L_{\rm
V} \sim 3.35$) and $\sim 4.2 \times 10^{7}~M_{\odot}$ for the
giant-dominated IMF ($M/L_{\rm V} \sim 12.3$).
Note that the large mass-to-light ratios for the giant-dominated IMF
are due to the larger number of massive stellar remnants (see
Figs. 16-17 in Maraston 1998). Taking also into account evaporation, the mass
can be decreased by an additional 1\% (Binney \& Tremaine 1987).

Therefore the mass of the evolved W3 cluster will be at least ~$\sim$
10 times larger than the mass of ${\omega}$ Cen
($2.9\times10^{6}~M_{\odot}$, Harris 1996), which is the most massive
cluster of the Milky Way.  As already suggested by Schweizer \&
Seitzer (1998), a dynamical estimate of the mass of W3, independent of
population synthesis, is required to confirm these estimates. This is
the subject of an ongoing project.

\subsection{Will NGC 7252 become a normal elliptical ?}

In the framework of hierarchical structure formation models (White \&
Rees 1978; Kauffmann, White \& Guiderdoni 1993; Cole et al. 1994),
elliptical galaxies are built up by merging disk systems.  It is
therefore interesting to know whether a merger-remnant like NGC 7252
will evolve into an object similar to a present day elliptical.  A
strong argument favouring this evolution is the observed light
distribution in the inner part of NGC 7252, which closely resembles
the $r^{1/4}$ de Vaucoulers light profile of elliptical galaxies
(Schweizer 1982). In spite of the fact that it is a recent merger, NGC
7252 already shows no sign of any major stellar disk component in its
inner part.  The colours of NGC 7252 will be close to the observed
range of normal ellipticals after a time of 1 to 4 Gyr from the
present, depending on when the star formation ceases (Fritze-von
Alvensleben \& Gerhard 1994b).  Yet there are two pieces of evidence
against the evolution of NGC 7252 to a bona fide elliptical.

Elliptical galaxies have $\alpha$-enhanced average stellar populations
(Worthey et al. 1992). This is found for ellipticals in dense
environments (Mehlert et al. 2000; Kuntschner 2000) and in the field
(Gonz{\'{a}}lez 1993; Trager et al. 2000; Beuing, Mendes de Oliveira
\& Bender 2000). Due to its very isolated position, NGC 7252 would
become a field elliptical, likely an $L^{*}$ object.  In Section 5 we
showed that neither the cluster population nor the diffuse light of
NGC 7252 show any sign of overabundance in
$\alpha$-elements. Therefore, the present average stellar population
of NGC 7252 cannot evolve into an average $\alpha$-enhanced stellar
population, in contrast to the observations of present day ellipticals
of the same luminosity.  Note that abundance ratios provide a stronger
constraint than colours because the latter can become sufficiently red
if enough time for passive evolution is allowed after star formation
stops.  The abundance ratios of Fe and Mg instead are fixed when star
formation stops and they cannot further change with passive evolution.
Note that the convective mixing (dredge-up) during the red-giant phase
does not alter the surface [Mg/Fe] abundance ratio, because these
elements are not synthetized in the interiors and therefore not
dredged-up (see e.g. Renzini \& Voli 1981).

Another complication is the number of metal-rich globular clusters
normalized to the galaxy light. It requires that a large number of new
globular clusters have to form during the merger process (e.g.~Forbes,
Brodie \& Grillmair 1997, Kissler-Patig et al.~1999).  We see from
this work and from the work of others mentioned in the introduction
that new globulars indeed form during a merger-induced starburst. It
is not, however, clear whether the number of new clusters is
sufficient to be compatible with the number of red clusters observed
in elliptical galaxies (see e.g. Harris 2000 for a review). As the
efficiency in cluster production obviously increases with the gas
content of the mergers (e.g. Kissler-Patig, Forbes \& Minniti 1998),
gas-rich mergers are better candidates. Gas-rich mergers, on the other
hand do not produce $\alpha$-enhanced stellar populations, as
discussed above. Furthermore the very young ages of the newly formed
clusters (Section 4) imply a large age difference compared to the old
cluster population belonging to the progenitor spirals.  While a large
number of studies have discovered bimodal GC colour and metallicity
distributions of globular clusters in more than half of the early-type
galaxies studied (Zepf \& Ashman 1993; Gebhardt \& Kissler-Patig
1999), spectroscopic and photometric investigations indicate that both
populations are old (Cohen et al. 1998; Kissler-Patig et al. 1998;
Kissler-Patig, Forbes \& Minniti 1998; Kundu et al. 1999; Puzia et
al. 1999).  Ellipticals with GCs populations that are well-separated
($>$ 8 Gyr) in age seem to be rare. NGC 7252 will therefore remain an
abnormal object for the present epoch even if, after the merger, it
resembles an elliptical in several respects.

\section{Summary and conclusions}

Recent studies have shown evidence for a population of bright, young
star clusters in the merger remnant NGC 7252 in which the AGB phase
transition should be detectable.  With this aim we obtained $K$
photometry for these clusters.  This is a new approach, previous
investigations have been confined to optical bands.  Indeed, in
intermediate-age populations ($0.2 {\rm Gyr} \la t \la 1$~Gyr) the AGB phase
is the major contributor to the total energy.  To analyze the data we
used models for simple stellar populations in which the contribution
of the AGB phase to the total light is calibrated on the
intermediate-age globular clusters of the Magellanic Clouds. 

\begin{itemize}
 \item{
The V$-$K vs. B$-$V diagram predicted by our models describes the observed
colours remarkably well and clearly shows the ongoing AGB phase
transition among these clusters.  This is the first detection of the
AGB phase transition outside the Local Group.}

\item{
The photometric ages derived from B$-$V and V$-$K are in excellent
agreement and are consistent with a metallicity $0.5\,$--$\,1~Z_{\odot}$ for
the studied clusters.} 

\item{ Most of the clusters appear to be $300\,$--$\,500$~Myr old,
consistent with previous age determinations based on optical
spectroscopy.}

\item{ One cluster (object W32) stands out as being $1\,$--$\,2$~Gyr
old. If this object belongs to the same family of star clusters
produced during the merger, then it must have formed before the first
pericentral contact and cluster (and star) formation in a gas-rich
disk-disk merger lasted at least $\,1$--$\,2$~Gyr.}

\item{
We further analyze the spectral features of the best observed cluster
W3 and of the diffuse light of the nucleus of NGC 7252, using
spectroscopy from the literature (Schweizer \& Seitzer 1998;
F. Schweizer, {\it private communication}). As expected from
simulations of the chemical evolution of merging spirals (Thomas,
Greggio \& Bender 1999), neither stars nor clusters show an overabundance in
$\alpha$-elements with respect to the iron group, in contrast to what
is observed in elliptical galaxies of the same luminosity. This
suggests that a present day disk-disk merger like NGC 7252 does not
evolve into a present day elliptical. Our results disfavour an
IMF substantially flatter than Salpeter during the merger-induced star
formation, because this produces $\alpha$-enhanced stellar populations.
}

\end{itemize}

\begin{acknowledgements}

We are grateful to Bryan Miller for providing us with an electronic
list of the UV and optical photometry, including unpublished data. Thanks also
to Francois Schweizer for providing us with his spectra in electronic form. 
Santi Cassisi is acknowledged for the stellar tracks and
isochrones used to compute the SSP models. We thank Daniel Thomas and
Paola Marigo for helpful discussions. Finally, we are grateful to
the anonymous referee for her/his relevant comments.
Data presented herein were
obtained at the W.M. Keck Observatory, which is operated as a
scientific partnership among the California Institute of Technology,
the University of California and the National Aeronautics and Space
Administration.  The Observatory was made possible by the generous
financial support of the W.M. Keck Foundation. This work was supported
in part by National Science Foundation grant number AST 9900732.  CM
is supported by the "Sonderforschungsbereich 375-95 f\"ur
Astro-Teilchenphysik" of the Deutsche Forschungsgemeinschaft.

\end{acknowledgements}

\end{document}